\documentclass[journal,twoside,web]{ieeecolor}
\usepackage{tmi_re}
\usepackage{cite}
\usepackage{amsmath,amssymb,amsfonts}
\usepackage{algorithmic}
\usepackage{graphicx}
\usepackage{textcomp}
\usepackage{graphics}
\usepackage{multirow}
\usepackage{subfigure}
\usepackage{soul}
\usepackage[table]{xcolor}
\usepackage{color, xcolor}
\soulregister{\cite}7 %
\soulregister{\ref}7 %
\soulregister{\eqref}7 %
\soulregister{\pageref}7 %
 
\usepackage{rotating}

\usepackage{hyperref}
\hypersetup{colorlinks=true, linkcolor=red, filecolor=red, urlcolor=red, citecolor=red}

\def\BibTeX{{\rm B\kern-.05em{\sc i\kern-.025em b}\kern-.08em
    T\kern-.1667em\lower.7ex\hbox{E}\kern-.125emX}}

\makeatletter
\def\fnum@figure{\textcolor{subsectioncolor}{\sf Fig.~\thefigure}}
\def\fnum@table{\textcolor{subsectioncolor}{\sf TABLE~\thetable}}
\makeatother

\begin{document}
\title{Cross-Site Severity Assessment of COVID-19 from CT Images via Domain Adaptation}
\author{Geng-Xin Xu, Chen Liu, Jun Liu, Zhongxiang Ding, Feng Shi, Man Guo, Wei Zhao, Xiaoming Li, Ying Wei, Yaozong Gao, Chuan-Xian Ren, Dinggang Shen, \IEEEmembership{Fellow, IEEE}
\thanks{This work was supported in part by the National Natural Science Foundation of China under Grant 61976229 and 81871337, in part by the National Key Research and Development Program of China under Grant 2018YFC0116400, in part by Chongqing Science and Health Joint Medical Research Project under Grant 2021MSXM052, in part by the Major Science and Technology Projects of Chongqing City under Grant cstc2018jszx-cyztzxX0017, in part by the Key Emergency Project of Pneumonia Epidemic of Novel Coronavirus Infection under Grant 2020SK3006, in part by the Emergency Project of Prevention and Control for COVID-19 of Central South University under Grant 160260005, in part by the Foundation from Changsha Scientific and Technical Bureau under Grant KQ2001001 and KQ1801115, in part by the Hunan Provincial Natural Science Foundation of China under Grant 2021JJ40895, in part by the Science and Technology Innovation Program of Hunan Province under Grant 2020SK53423, and in part by the Clinical Research Center For Medical Imaging In Hunan Province under Grant 2020SK4001. (Geng-Xin Xu, Chen Liu, Jun Liu, Zhongxiang Ding, and Feng Shi contributed equally to this work.) (Corresponding authors: Chuan-Xian Ren; Dinggang Shen.)}
\thanks{Geng-Xin Xu is with the School of Mathematics, Sun Yat-sen University, Guangzhou 510275, China.}
\thanks{Chen Liu, Man Guo, and Xiaoming Li are with the Department of Radiology, Southwest Hospital, Third Military Medical University (Army Medical University), Chongqing 400038, China.}
\thanks{Jun Liu and Wei Zhao are with the Department of Radiology, the Second Xiangya Hospital, Central South University, Changsha 410011, China. Jun Liu is also with the Department of Radiology Quality Control Center, Hunan Province, Changsha 410011, China.}
\thanks{Zhongxiang Ding is with the Department of Radiology, Hangzhou First People’s Hospital, Zhejiang University School of Medicine, Hangzhou 310027, China.}
\thanks{Feng Shi, Ying Wei, and Yaozong Gao are with the Department of Research and Development, Shanghai United Imaging Intelligence Co., Ltd., Shanghai 200232, China.}
\thanks{Chuan-Xian Ren is with the School of Mathematics, Sun Yat-sen University, Guangzhou 510275, China, also with Pazhou Lab, Guangzhou 510330, China, and also with the Key Laboratory of Machine Intelligence and Advanced Computing (Sun Yat-sen University), Ministry of Education, Guangzhou 510275, China (e-mail: rchuanx@mail.sysu.edu.cn).}
\thanks{Dinggang Shen is with the Department of Research and Development, Shanghai United Imaging Intelligence Co., Ltd., Shanghai 200232, China, also with the School of Biomedical Engineering, ShanghaiTech University, Shanghai 201210, China, and also with the Department of Artificial Intelligence, Korea University, Seoul 02841, Republic of Korea (e-mail: dinggang.shen@gmail.com).}
}

\maketitle
\thispagestyle{empty}

\begin{abstract}
Early and accurate severity assessment of Coronavirus disease 2019 (COVID-19) based on computed tomography (CT) images offers a great help to the estimation of intensive care unit event and the clinical decision of treatment planning. To augment the labeled data and improve the generalization ability of the classification model, it is necessary to aggregate data from multiple sites. This task faces several challenges including class imbalance between mild and severe infections, domain distribution discrepancy between sites, and presence of heterogeneous features. In this paper, we propose a novel domain adaptation (DA) method with two components to address these problems. The first component is a stochastic class-balanced boosting sampling strategy that overcomes the imbalanced learning problem and {improves the classification performance on poorly-predicted classes}. The second component is a representation learning that guarantees three properties: 1) domain-transferability by prototype triplet loss, 2) discriminant by conditional maximum mean discrepancy loss, and 3) completeness by multi-view reconstruction loss. Particularly, we propose a domain translator and align the heterogeneous data to the estimated class prototypes {(i.e., class centers)} in a hyper-sphere manifold. Experiments on cross-site severity assessment of COVID-19 from CT images show that the proposed method can effectively tackle the imbalanced learning problem and outperform recent DA approaches.
\end{abstract}

\begin{figure}
\begin{center}
   \includegraphics[width=0.75\linewidth]{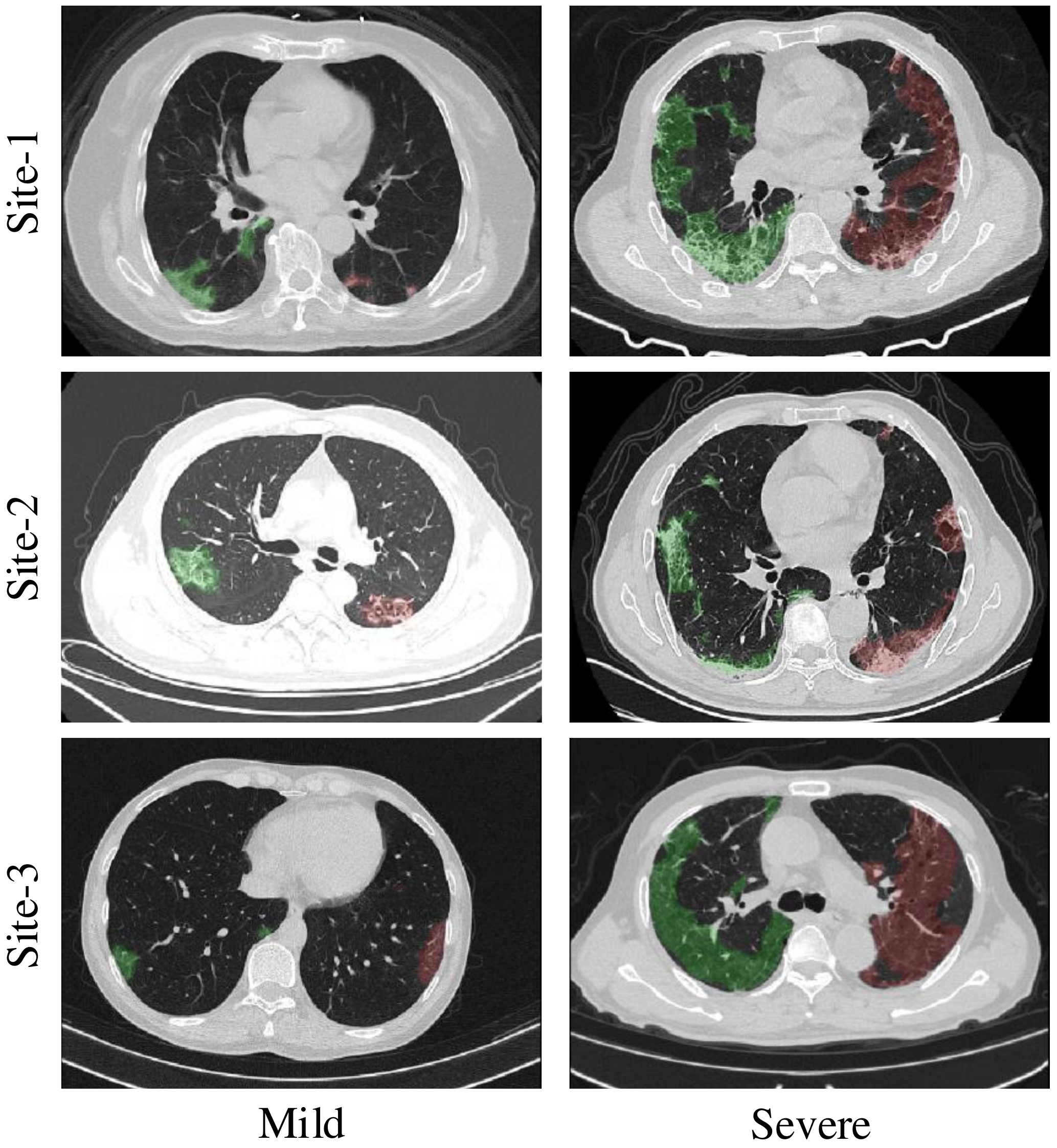} 
\end{center}
   \caption{Examples of chest CT images with infection of COVID-19 from different clinical sites. Top to bottom: three sites. Left: mild; Right: severe. It can be observed that visual discrepancies (on the appearance and contrast) exist between sites because of the difference in imaging conditions (e.g., imaging protocols and scanner vendors).}
\label{fig:Examples of CT images}
\end{figure}


\section{Introduction}
\label{sec:Introduction}

\IEEEPARstart{S}{ince} the end of 2019, Coronavirus disease 2019 (COVID-19) caused by the SARS-CoV-2 virus, has been widely spread around the world. The common manifestations of most infected patients are fever, dry cough, and malaise \cite{wang2020novel}. Some of these patients progress rapidly with acute respiratory distress syndrome, septic shock, and multiple organ failure, etc., resulting in permanent damage or even death \cite{huang2020clinical, li2020early, chen2020epidemiological}. Therefore, early and accurate severity assessment of COVID-19 is of crucial importance for the estimation of intensive care unit admission, oxygen therapy, and timely treatment.

Since most COVID-19 patients have some typical radiographic features in chest computed tomography (CT), e.g., ground-glass opacity and pulmonary consolidation, chest CT plays a key role in the diagnosis and quantification of COVID-19 \cite{song2021deep, yazdani2020covid, ko2020covid, saeedizadeh2021covid, kang2020diagnosis, chassagnon2020ai, ouyang2020dual, minaee2020deep, chaganti2020automated}. For instance, Song {\textit{et al.}} \cite{song2021deep} propose a deep learning-based CT diagnosis system, which aims at assisting doctors to detect the patients with COVID-19 and automatically localizing the ground-glass opacity in CT images. In \cite{yazdani2020covid}, Yazdani {\textit{et al.}} propose a deep learning framework based on the attentional convolutional network and reach an accuracy of 92\% for COVID-19 prediction from chest CT images. Chassagnon {\textit{et al.}} \cite{chassagnon2020ai} design a deep learning-based pipeline for COVID-19 disease quantification. Moreover, Tang {\textit{et al.}} \cite{tang2021severity} extract a series of quantitative features from chest CT images and train a random forest model to assess the severity.

Under the regime of limited CT imaging data with single site (hospital) coverage, the aforementioned models may learn underdetermined parameters and generalize poorly. To alleviate this problem, introducing more data from other sites is an alternative strategy. Such a cross-site data transfer strategy has been verified in other medical image analyses, e.g., prostate segmentation \cite{gibson2018inter, liu2020ms}, pneumonia detection \cite{zech2018variable}, and COVID-19 CT diagnosis \cite{wang2020contrastive}. However, to the best of our knowledge, no previous study has proposed a method for the cross-site severity assessment of COVID-19. There may be several complicated challenges in this task. Firstly, the number of training samples for each class is imbalanced. For example, according to the report from the WHO-China Joint Mission on COVID-19, around 20\% developed severe to critical disease (referred to as severe in this work) and the rest remained mild to moderate (referred to as mild in this work) out of 55,924 patients \cite{verity2020estimates}. Secondly, as is shown in Fig. \ref{fig:Examples of CT images}, due to the difference in imaging conditions, the CT images from different cites show discrepancies on appearance and contrast. Directly aggregating the data may contribute little to downstream tasks and lead to negative transfer. Thirdly, most recent works adopt deep learning methods based on original medical images. This manner may have difficulty in exploiting the expert knowledge, which offers a great deal of severity-related features.

In this paper, we propose a novel perspective with domain adaptation (DA) method for the cross-site severity assessment of COVID-19 from CT images. It aims to transfer discriminant knowledge from a clinical site (source domain) to another site (target domain), where the source domain contains much labeled data whereas the target domain contains a little labeled data. Our method consists of two key components: a stochastic class-balanced boosting sampling (SCBS) phase and a representation learning phase. In SCBS, we stochastically update the training dataset according to sample imbalance rate and classification difficulty (defined as the self-information of prediction in Section \ref{subsec:Imbalanced Learning}) of classes, which aims at fair classification across major versus minor classes, and easy versus hard (i.e., well-predicted versus poorly-predicted by the model) classes. Then, we propose a domain translator to project the source domain data into a latent space while preserving the local structure, which is optimized by a well-designed prototype triplet loss. To strengthen the discriminant and guarantee the completeness of the latent space, we adopt a conditional maximum mean discrepancy (CMMD) loss and a multi-view reconstruction loss on the representation. We refer to our method as Fair cross-domain adaptation with LAtent REpresentations (FLARE).

We summarize the contributions of this paper as follows.
\begin{itemize}
\item We propose a DA method, FLARE, for the cross-site diagnosis of COVID-19 severity from CT images. FLARE can deal with the class-imbalance problem in tandem with the domain shift problem between multi-site datasets.
\item We propose a domain translator to alleviate the negative transfer problem. In addition, we learn a latent space possessing three properties: domain-transferability, discriminant, and completeness.
\item We extend the FLARE method to the multi-source DA scenario. In particular, we design multiple domain translators and weighted classifiers to promote the performance of the DA method.
\item To evaluate the effectiveness of FLARE, we collect CT images from multiple clinical sites and assess COVID-19 severity. Extensive experiment results show that FLARE outperforms recent DA and imbalanced learning methods.
\end{itemize}

The rest of this paper is organized as follows. In Section \ref{sec:Related Works}, related works on COVID-19 diagnosis and quantification, DA method, and imbalanced learning strategy are briefly reviewed. In Section \ref{sec:Methods}, the proposed FLARE is presented. Experimental validation against recent methods is conducted in Section \ref{sec:Experiments}. Analysis and discussion on our work are described in Section \ref{sec:Discussion}. Conclusions are given in Section \ref{sec:Conclusion}.

\section{Related Works}
\label{sec:Related Works}

\subsection{COVID-19 Diagnosis and Quantification}
\label{subsec:COVID-19}

In the past few months, numerous methods have been proposed for the diagnosis and quantification of COVID-19 \cite{kang2020diagnosis, ouyang2020dual, song2021deep, yazdani2020covid, ko2020covid, saeedizadeh2021covid, li2020automated}. In particular, Ko {\textit{et al.}} \cite{ko2020covid} develop a deep learning framework to diagnose COVID-19 pneumonia, and obtain 99.87\% accuracy in detecting COVID-19 from institutional data and 96.97\% accuracy in detecting COVID-19 from external validation data. In \cite{saeedizadeh2021covid}, Saeedizadeh {\textit{et al.}} present a deep learning framework for COVID-19 segmentation from CT images, where a novel connectivity-promoting regularization term is employed to impose desired connectivity requirements. Moreover, Kang {\textit{et al.}} \cite{kang2020diagnosis} extract a series of features from CT images and propose a structured latent multi-view representation learning for COVID-19 diagnosis. Xi {\textit{et al.}} \cite{ouyang2020dual} develop a dual-sampling strategy and online attention module for COVID-19 diagnosis. In addition, Zhu {\textit{et al.}} \cite{zhu2020joint} predict the probability and the time of COVID-19 developing severe symptoms using a joint classification and regression method based on chest CT scan data. To effectively combine CT data from different sites, Wang {\textit{et al.}} \cite{wang2020contrastive} use a contrastive learning method to enhance domain-invariance of representations. Li {\textit{et al.}} \cite{li2020automated} use a convolutional Siamese neural network-based method to predict the radiographic pulmonary disease severity from chest radiographs. These approaches may learn underdetermined parameters when there is little labeled data in a single site. We address this problem via a cross-site data transfer strategy in this paper.

\subsection{Domain Adaptation Method}
\label{subsec:DA}

DA methods aim to find a latent space for source and target domains, so that the discrimination information in the source domain can be efficiently transferred to the recognition task in the target domain \cite{pan2010survey, dong2020weakly, li2020enhanced, ren2020learning}. Generally, DA methods can be divided into three categories: unsupervised DA methods, weakly-supervised DA methods, and semi-supervised DA methods. Classical unsupervised DA methods achieve domain alignment by moment matching \cite{long2015learning} or adversarial training strategies \cite{ganin2016domain, long2018conditional}. In addition, optimal experimental design \cite{ren2019heterogeneous} and manifold propagation \cite{luo2020unsupervised} make a prominent breakthrough in adaptive feature learning. Weakly-supervised DA methods aim to tackle the problem where the target domain has some weakly-labeled data. For example, Dong {\textit{et al.}} \cite{dong2020weakly} assume that there are some image-level labels but pixel-level ones on the target domain, and propose a quantified transferability mechanism for endoscopic lesions segmentation. Semi-supervised DA methods assume that there is a little labeled data on the target domain. {When the domain shift is large, semi-supervised DA methods can obtain better results compared with unsupervised DA methods, due to the use of partial supervision information in the target domain \cite{li2018semi, chen2019a}.} For instance, the domain adaptation by covariance matching (DACoM) method \cite{li2018semi} matches the distributions by minimizing the covariance in two domains. However, these methods lack consideration of variation in the target domain. Recently, Saito {\textit{et al.}} \cite{saito2019semi} propose a novel minimax entropy (MME) approach, which gradually moves the prototypes closer to unlabeled target data. Tseng {\textit{et al.}} \cite{Tseng2020Cross-Domain} design a transformation layer on the top of the feature extractor to simulate various domain feature distributions. Motivated by these works, we propose a domain translator to project the source data around the target prototypes. This manner simulates the target data that are difficult to classify, and enables the model to learn discriminative class boundaries on the target domain.

\subsection{Imbalanced Learning Strategy}
\label{subsec:Imbalanced Learning}

Imbalanced data is fairly common in the real world, especially in the field of medical imaging. There are two widely used strategies to tackle this problem, i.e., sampling strategy and cost-sensitive learning strategy. The sampling strategy tries to construct a balanced sample set or feature space \cite{wang2021towards, chawla2002smote, han2005borderline, he2008adasyn, batista2004study}, and the cost-sensitive learning strategy weights the loss function using the well-designed cost \cite{lin2017focal}. Most aforementioned approaches aim to learn the classifier jointly with the feature extractor. However, the mechanism of imbalanced learning is unclear. Recently, Zhou {\textit{et al.}} \cite{zhou2020bbn} discover that the key to achieving a satisfactory long-tailed recognition ability is the classifier learning. Similarly, a new approach named classifier re-training (cRT) \cite{Kang2020Decoupling} first jointly learns the feature extractor and the classifier based on the raw dataset, and then fine-tunes the classifier on a balanced sample set. In addition to the class imbalance, the classification difficulty of the minor class also attracts increasing attention. For instance, Lin {\textit{et al.}} \cite{lin2017focal} modify the cross-entropy loss to a focal loss by down-weighting the loss for well-classified sample, which significantly improves the speed and accuracy of the object detector. Yang {\textit{et al.}} \cite{yang2019self} use class size and recognition difficulty to define the complexity level of a class, and then propose a corresponding curriculum reconstruction (CR) strategy for clinical skin disease recognition.

More recently, some works \cite{minaee2020deep, khalifa2020detection, apostolopoulos2020covid, maghdid2021diagnosing} investigate the transfer learning methods and achieve favorable performance for COVID-19 diagnosis. The differences between these studies and our work are three-fold: 
1) The task. Most recent methods address the problems in discriminating between COVID-19 and other community-acquired pneumonia or normal. However, we propose a pipeline for the severity assessment of COVID-19, which is by its nature heavily imbalanced; 
2) The manner of using transfer learning framework. Some previous works adopt the popular convolutional neural networks pre-trained on a large-scale image dataset (e.g., ImageNet), and fine-tune them for COVID-19 recognition task. In this paper, we focus on learning domain-transferable and discriminative features from one site to another one. This goal helps us to introduce more labeled data from other sites and achieve data-augmentation; 
3) The type of used features. Some related works train models based on a single type of feature, whereas we exploit multiple types of heterogeneous features. This helps us to take more effective information into account, especially the features extracted with expert knowledge.

\section{Methods}
\label{sec:Methods}

\begin{figure*}
\begin{center}
\includegraphics[width=1\linewidth]{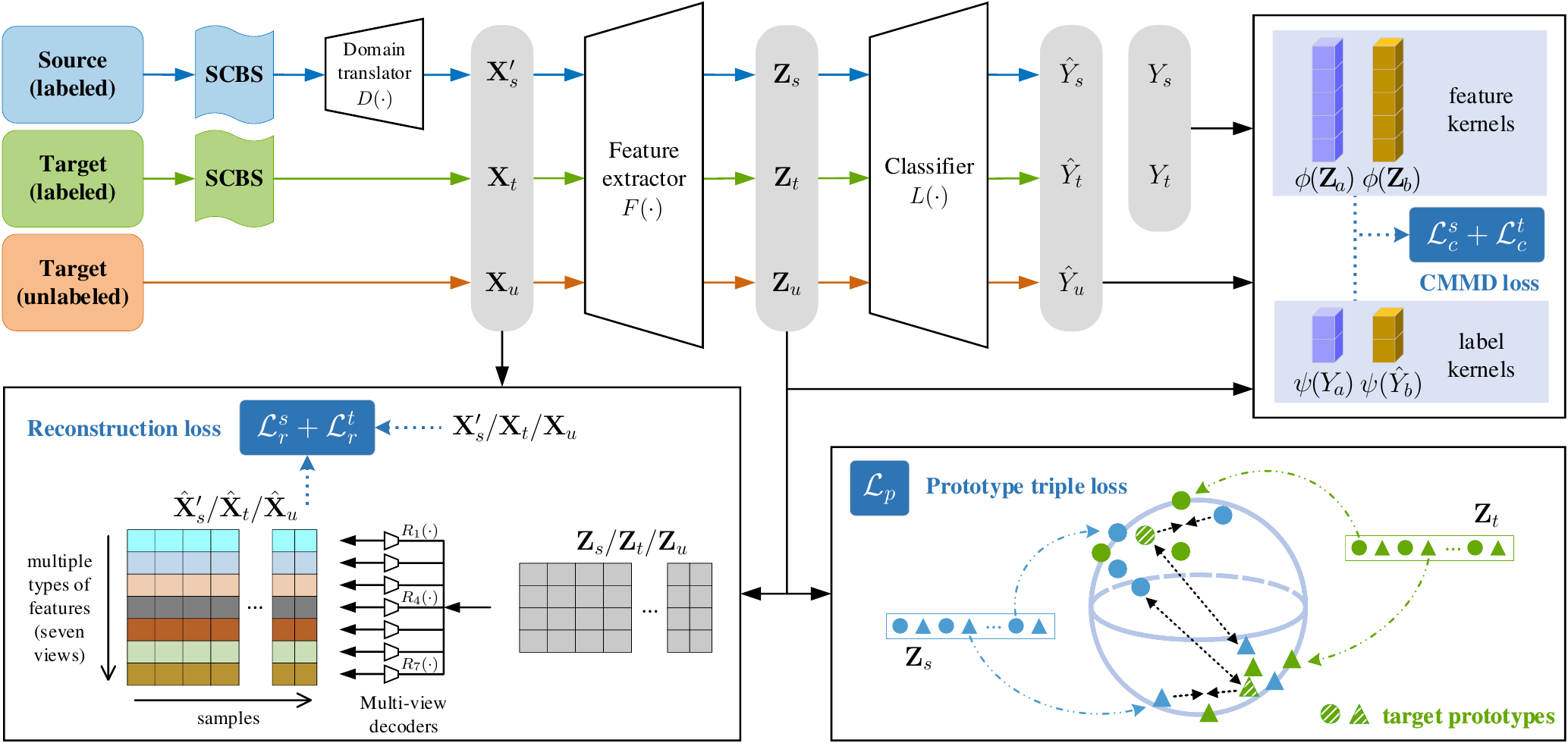} 
\end{center}
   \caption{Flowchart of the FLARE method. It updates the class distribution of the samples in the source domain and target domain by an SCBS strategy. After optimizing a specially designed domain translator on the source data, FLARE learns a latent space with three properties: (a) domain-invariance by the prototype triplet loss $\mathcal{L}_p$, (b) discriminant by the conditional maximum mean discrepancy (CMMD) loss ($\mathcal{L}_c^s + \mathcal{L}_c^t$), and (c) completeness by the multi-view reconstruction loss ($\mathcal{L}_r^s + \mathcal{L}_r^t$).}
\label{fig:Overall architecture of the FLARE}
\end{figure*}

An overview of the proposed FLARE method is shown in Fig. \ref{fig:Overall architecture of the FLARE}. It concentrates on imbalanced learning and representation learning simultaneously. With regard to imbalanced learning, an SCBS strategy is designed to generate a balanced data distribution. To reduce the domain distribution discrepancy, we propose a domain translator on the source domain. Subsequently, FLARE learns a latent space possessing three properties: 1) domain-invariance by the prototype triplet loss, 2) discriminant by the CMMD loss, and 3) completeness by the multi-view reconstruction loss.

We formulate the framework of FLARE as follows. It includes a domain translator $D(\cdot)$, a feature extractor $F(\cdot)$, a classifier $L(\cdot)$, and $V$ parallel reconstruction networks $\left \{ R_v(\cdot) \right \}_{i=1}^{V}$ ($V=7$ here). All these modules adopt fully connected neural networks (FCNs), and share weights between domains, except that the domain translator $D(\cdot)$ is applied only to the source domain. Thus, $F \circ D (\cdot)$ projects source data onto the latent space, and $F (\cdot)$ projects target data onto the same space. In this latent space, features from two domains are aligned in a hyper-sphere manifold. Finally, the extracted features will be fed into two pipelines. One is the classifier $L(\cdot)$ for the classification task. Another pipeline is the parallel networks $\left \{ R_v(\cdot) \right \}_{i=1}^{V}$ for reconstructing multi-view features.

\subsection{Stochastic Class-balanced Boosting Sampling (SCBS) Strategy}
\label{subsec:Stochastic Class-balanced Boosting Sampling}

We propose an SCBS strategy to overcome the imbalanced learning problem and improve the classification performance on poorly-predicted classes.

Consider a given training dataset with $C$ classes $S=\cup_{j=1}^C S_j$, where $S_j$ denotes the $j$-th class data subset, $\left | S_1 \right | \geq \left | S_2 \right | \geq \cdots \geq \left | S_C \right |$, and $\cap_{j=1}^C S_j = \emptyset$. In most sampling strategies, the probability of sampling a sample from class $j$ is defined as
\begin{equation}
\label{equ:sampling strategies}
r^q(j) = {\left | S_j \right |}^q / \sum\nolimits_{k=1}^{C} {\left | S_k \right |}^q,
\end{equation}
where $q \in \left [ 0,1 \right ]$. When $q=1$, samples generated by \eqref{equ:sampling strategies} are subject to the class distribution as the original dataset. When $q=0$, \eqref{equ:sampling strategies} is the popular class-balanced sampling strategy, for example, random oversampling (by generating $\left | S_1 \right |$ samples from the minor class subset $S_k$ ($k=2, \cdots, C$)).

We define the classification difficulty for class $j$, i.e.,
\begin{equation}
\label{equ:related classification difficulty}
r^{DIFF}(j) = \mathbb{E}_{i \in S_j} I \left [ p(y_i | \mathbf{x}_i) \right ] / \sum\nolimits_{k=1}^{C} \left \{ \mathbb{E}_{i \in S_k} I \left [ p(y_i | \mathbf{x}_i) \right ] \right \},
\end{equation}
where $I \left [ p(y_i | \mathbf{x}_i) \right ] = - \log\left [ p(y_i | \mathbf{x}_i) \right ]$ is the self-information to measure the classification difficulty of an instance $\mathbf{x}_i$ with label $y_i$. The denominator in \eqref{equ:related classification difficulty} is a normalization constant.

Let $T$ be the total number of epochs. At the $t$-th epoch, the probability of sampling a sample from class $j$ in our SCBS is
\begin{equation}
\label{equ:SCBS}
r^{SCBS}_t(j) =
\begin{cases}
(1-w_t) r^0(j) + w_t r^{DIFF}(j), & \varepsilon_t \leq \delta,\\
r^{SCBS}_{t-1}(j),            & \text{otherwise},
\end{cases}
\end{equation}
where $r^{SCBS}_0(j) = r^1(j)$, $\varepsilon_t$ is a random sample from a uniform distribution, and $w_t$ is a trade-off parameter that controls the relative importance of poorly-predicted classes mining. Threshold $\delta \in ( 0, 1 ]$ reflects the frequency of updating class distribution in SCBS, and it is set to be $0.3$ in the experiments. At the beginning of the training process, the predictions of some samples are unreliable, and the classification difficulty defined by \eqref{equ:related classification difficulty} cannot detect the poorly-predicted classes well. Hence, we let $w_t$ be a relatively small value and focus on addressing the imbalanced learning problem when $t$ is small. Along with the optimization process (the decrease of the objective function), the predictions of most training samples become reliable. Then, we use a larger value for $w_t$ and focus on addressing the poorly-predicted classes mining problem. With the above analysis, $w_t$ can be calculated as a cosine-based function $w_t = \left [ 1 - \cos (t \pi / T) \right ] / 2$. In our experiments, we randomly generate a sample $\varepsilon_t$ from a uniform distribution at the end of each epoch. If $\varepsilon_t$ is larger than $\delta$, $r^{SCBS}_t(j)$ remains unchanged; otherwise, we compute the expectation in \eqref{equ:related classification difficulty} by averaging over all training samples with class $j$, and then update $r^{SCBS}_t(j)$ by \eqref{equ:SCBS}.

Our SCBS has two advantages as follows. Firstly, since the minor class has few samples, a single oversampling operation may possess a single data distribution, and thus the model will over-fit the corresponding distribution. Instead, SCBS stochastically implements the sampling strategy during the training phase. This manner simulates the data variation and prevents model over-fitting. Secondly, at the beginning of the training phase, SCBS focuses on the imbalanced learning problem. Along with the increase of $w_t$, it pays more attention to the poorly-predicted classes mining problem and further boosts the recognition performance.

\subsection{Domain Adaptation via the Prototype Triplet Loss}
\label{subsec:DA via Prototype triplet loss}

The COVID-19 CT images present two characteristics as follows. On one hand, within a single site, the samples are usually with large variations. For example, lung deterioration varies considerably between patients because of coexisting diseases. Thus, when single-site data is insufficient, it is difficult for a model only optimized on few label samples to achieve excellent generalization ability. On the other hand, the domain shift is large as shown in Fig. \ref{fig:Examples of CT images}. Directly introducing medical images from other sites may lead to negative transfer. Hence, learning domain-adaptive features is important for the task of cross-site COVID-19 severity assessment.

For a semi-supervised DA task, we have a source domain $\mathcal{D}_s = \left \{ (\mathbf{x}_i^s, y_i^s) \right \}_{i=1}^{n_s}$, a labeled target domain $\mathcal{D}_t = \left \{ (\mathbf{x}_i^t, y_i^t) \right \}_{i=1}^{n_t}$, and an unlabeled target domain $\mathcal{D}_u = \left \{ (\mathbf{x}_i^u) \right \}_{i=1}^{n_u}$, where $y_i \in \left \{ 1, \cdots, C \right \}$. Our goal is to train a model on $\mathcal{D}_s$, $\mathcal{D}_t$, and $\mathcal{D}_u$, and then evaluate on $\mathcal{D}_u$. Particularly, attributed to the transferred discrimination knowledge from $\mathcal{D}_s$, the learned model is expected to outperform the target-only model which is trained only on $\mathcal{D}_t$.

To transfer discriminant information from the source domain to the target domain, we design a prototype triplet loss as
\begin{equation}
\label{equ:prototype triplet loss}
\begin{aligned}
\mathcal{L}_p = \frac{1}{C} \sum\nolimits_{j=1}^{C} \left \{  \left [ \max d \left ( \mathbf{p}_j, F \circ D (\mathbf{x}^{s_{j,n}}) \right )
                                                                                                              \right.
                             \right.\\ 
                             \left.
                                                                                                              \left.
                                                                                                                    -\min d \left ( \mathbf{p}_j, F \circ D (\mathbf{x}^{s_{j,p}}) \right )
                                                                                                              \right ] + \alpha
                             \right \} _{+},\\
\end{aligned}
\end{equation}
where the prototype $\mathbf{p}_j = \mathbb{E}_{\mathbf{x}_i^{t} \in \mathcal{D}_t, y_i^t=j} F (\mathbf{x}_i^{t})$ is the class center, $\mathbf{x}^{s_{j,p}}$ is a source sample with class $j$, $\mathbf{x}^{s_{j,n}}$ is a source sample with class different from $j$, $d \left ( \cdot, \cdot \right )$ is the cosine similarity measure, and $\alpha \in [0,2]$. To construct a separable space, we normalize all features into a hyper-sphere manifold. {In this hyper-sphere, we estimate the prototype $\mathbf{p}_j$ for each class on the target domain and update $\mathbf{p}_j$ by averaging the normalized features of all target labeled samples with class $j$ at the end of each training epoch. Then, we project the source samples near their corresponding prototypes in the latent space, which preserves the discriminative local structure of the source domain}. Moreover, in \eqref{equ:prototype triplet loss}, we mainly focus on the data that is difficult to classify, i.e., the hard positive and negative examples. Specifically, for a target prototype $\mathbf{p}_j$, we push the closest negative example $\mathbf{x}^{s_{j,n}}$ farther to it and pull the farthest positive example $\mathbf{x}^{s_{j,p}}$ closer to it. In this process, the local structure (or the relationship of data) in the source domain is preserved, i.e., the intra-class similarity of latent features is larger than the inter-class one. Thus, minimizing \eqref{equ:prototype triplet loss} helps to transfer discriminant knowledge between domains.

Different from those DA methods via moment matching or adversarial training manner, the proposed prototype triplet loss guarantees domain alignment as well as data-augmentation. As the hyper-sphere in Fig. \ref{fig:Overall architecture of the FLARE} shows, the source samples are projected around the estimated target prototypes. This manner of data transformation can simulate various data distributions in the target domain, especially those samples that are far away from the prototypes with the same class, but close to the prototypes with different classes. Thus, FLARE has stronger generalization ability and superior performance for successfully classifying the unlabeled target data.

With regard to the triplet loss in \eqref{equ:prototype triplet loss}, there are many other ranking loss functions for the task of metric learning. One of the typical works is the contrastive loss \cite{hadsell2006dimensionality}, which aims to maximize all positive similarities (between intra-class samples) and minimize all negative ones (between inter-class samples). Compared with the contrastive loss, the advantages of the prototype triplet loss are two-fold. 1) The contrastive loss maximizes all positive similarities and minimizes all negative similarities separately, whereas the prototype triplet loss maximizes the difference between positive similarity and negative similarity with margin $\alpha$. Since the relationship of two similarities is taken into account, the feature learned via the prototype triplet loss has better class-separating ability, as shown in Section \ref{subsec:Visualization}. 2) The contrastive loss encourages all intra-class distances to approach 0. However, the prototype triplet loss merely tries to keep all intra-class distances smaller than any inter-class one above a certain threshold. This is a less restrictive objective and can lead to better performance on the target data, as shown in Section \ref{subsec:Ablation Study}.

\subsection{Discriminant Learning via the CMMD Loss}
\label{subsec:CMMD}

Due to the difference in imaging conditions and patient demographics, the COVID-19 CT images show large appearance discrepancies. These discrepancies (in the original input space) will hold up the training process and affect the prediction results. Thus, learning discriminative and compact features is important for the severity classification task.

Kernel-based pattern analysis has been validated in improving classification performance~\cite{ren2019learning}. To make use of the kernel method for classification, one can directly map the input data onto a reproducing kernel Hilbert space (RKHS) and establish a classifier followed by a cross-entropy (CE) loss. However, this manner relies on the selection of a suitable kernel function, which is expected to effectively capture the difference between intra-class similarity and inter-class one. To address this problem, Ren {\textit{et al.}}~\cite{ren2019learning} propose a new kernel learning method for CMMD-based image classification. Motivated by its effectiveness in learning a representative kernel function through placing the CMMD loss in the representation learning phase, we design a CMMD loss in FLARE as follows.

Suppose $\mathcal{F}$ and $\mathcal{G}$ are the RKHS corresponding to an input $X$ and its true label $Y$, respectively. CMMD intuitively characterizes the discrepancy between two conditional distributions $P(Y|X)$ and $P(\hat{Y} | X)$, where $\hat{Y}$ represents the predicted label\footnote{Note that the core task in a classification problem is to train a predictive model so that $P(Y|X) = P(\hat{Y} | X)$ holds or approximately holds. Thus, CMMD is capable of solving the classification task.}. Given two batches of samples $\mathcal{D}_{XY}^a=\{ (\mathbf{x}_i^a, y_i^a) \}_{i=1}^n$ and $\mathcal{D}_{X\hat{Y}}^b=\{ (\mathbf{x}_i^b, \hat{y}_i^b) \}_{i=1}^n$, from $P(Y|X)$ and $P(\hat{Y} | X)$, respectively. The empirical estimation of CMMD is
\begin{equation}
\label{equ:general CMMD loss}
\mathcal{L}_{\mathrm{CMMD}} = \left \| \hat{C}_{Y | \mathbf{X}}^a - \hat{C}_{\hat{Y} | \mathbf{X}}^b \right \|_{\mathcal{F} \bigotimes \mathcal{G}}^2,
\end{equation}
where $\bigotimes$ denotes the tensor product operating, $\hat{C}_{Y | \mathbf{X}}^a = \Psi_a (K_a+ \tau \mathbf{I})^{-1} \Phi_a^T$ is the kernel embedding of conditional distribution, $\Psi_a = [\psi(y_1^a), \cdots, \psi(y_n^a)]$, $\Phi_a = [\phi(\mathbf{x}_1^a), \cdots, \phi(\mathbf{x}_n^a)]$, $K_a = \Phi_a^{T}\Phi_a$ \cite{baker1973joint}. Other variables with subscript $b$ are similarly defined on $\mathcal{D}_{X\hat{Y}}^b$. By minimizing \eqref{equ:general CMMD loss}, the difference between $P(Y|X)$ and $P(\hat{Y} | X)$ is reduced and then a predictive model is obtained.

The CMMD loss in FLARE consists of two terms on source and target domains respectively. As for the first term, we randomly sample two batches, $\mathcal{D}_s^a = \{ (\mathbf{x}_i^{s,a}, y_i^{s,a}) \}_{i=1}^{n_s}$ and $\mathcal{D}_s^b = \{ (\mathbf{x}_i^{s,b}, \hat{y}_i^{s,b}) \}_{i=1}^{n_s}$ from source domain $\mathcal{D}_s$ with $\hat{y}_i^{s,b} = L \circ F \circ D (\mathbf{x}_i^{s,b})$. Let $\mathbf{X}_{s}^a = [ \mathbf{x}_1^{s,a}, \cdots, \mathbf{x}_{n_{s}}^{s,a} ]$, $Y_{s}^a = [ y_1^{s,a}, \cdots, y_{n_{s}}^{s,a} ]$. Variables $\mathbf{X}_{s}^b$ and $\hat{Y}_{s}^b$ are defined in a similar way on the dataset $\mathcal{D}_{s}^b$. As~\cite{ren2019learning}, we apply the kernel functions on the latent features ($\mathbf{Z}_{s}^a$ and $\mathbf{Z}_{s}^b$) but the raw data ($\mathbf{X}_{s}^a$ and $\mathbf{X}_{s}^b$). The CMMD loss on the source domain is
\begin{align}
\label{equ:CMMD loss on the source domain}
\mathcal{L}_c^s = &\left \| \hat{C}_{Y_{s}^a | \mathbf{Z}_{s}^a} - \hat{C}_{\hat{Y}_{s}^b | \mathbf{Z}_{s}^b} \right \|_{\mathcal{F} \bigotimes \mathcal{G}}^2,
\end{align}
where $\mathbf{Z}_{s}^a = F \circ D (\mathbf{X}_{s}^a)$, $\mathbf{Z}_{s}^b = F \circ D (\mathbf{X}_{s}^b)$, and $\hat{Y}_{s}^b = L (\mathbf{Z}_{s}^b)$.

As for the second term, we first construct a complete target domain, i.e., $\mathcal{D}_{tu} = \mathcal{D}_t \cup \mathcal{D}_u$. Then, we randomly sample two batches, $\mathcal{D}_t^a = \{ (\mathbf{x}_i^{t,a}, y_i^{t,a}) \}_{i=1}^{n_t}$ and $\mathcal{D}_{tu}^b = \{ (\mathbf{x}_i^{tu,b}, \hat{y}_i^{tu,b}) \}_{i=1}^{n_t}$, from $\mathcal{D}_t$ and $\mathcal{D}_{tu}$, respectively. Correspondingly, we can define $\mathbf{X}_{t}^a$, $Y_{t}^a$, $\mathbf{X}_{tu}^b$, and $\hat{Y}_{tu}^b$ in a similar way as $\mathbf{X}_{s}^a$, $Y_{s}^a$, $\mathbf{X}_{s}^b$, and $\hat{Y}_{s}^b$. The CMMD loss on the target domain is formulated as
\begin{align}
\label{equ:CMMD loss on the target domain}
\mathcal{L}_c^t = &\left \| \hat{C}_{Y_{t}^a | \mathbf{Z}_{t}^a} - \hat{C}_{\hat{Y}_{tu}^b | \mathbf{Z}_{tu}^b} \right \|_{\mathcal{F} \bigotimes \mathcal{G}}^2,
\end{align}
where $\mathbf{Z}_{t}^a = F (\mathbf{X}_{t}^a)$, $\mathbf{Z}_{tu}^b = F (\mathbf{X}_{tu}^b)$, and $\hat{Y}_{tu}^b = L (\mathbf{Z}_{tu}^b)$. 
Equation \eqref{equ:CMMD loss on the target domain} presents a semi-supervised training manner on the target domain. Specifically, it simultaneously uses labeled and unlabeled data. As there may not be sufficient labeled data in each hospital in practice, making full use of the unlabeled data helps to learn more discriminative kernel functions, which is further verified by the experiment in Section \ref{subsec:Performance Under Various Training Regimes}.

{The differences between our FLARE and some related works are as follows. 1) In \cite{ren2019learning}, the CMMD loss is used on traditional classification tasks where training and testing data is subject to the same distribution, whereas we design a CMDD loss on the DA task. In addition, compared with \cite{ren2019learning}, we consider both supervised and semi-supervised learning scenarios in a framework. 2) Compared with some popular training manners (e.g., using a CE loss), FLARE places a CMMD loss on the latent space to learn more discriminative features. Specifically, in \eqref{equ:CMMD loss on the source domain} and \eqref{equ:CMMD loss on the target domain}, we adopt $F (\cdot)$ as the way of improved kernel learning on the source and target domains. This enables the learned kernel values to have better separating ability. Hence, the latent space in FLARE possesses well discriminant for the COVID-19 severity assessment.}

\subsection{Multi-view Representation Learning via the Reconstruction Loss}
\label{subsec:Reconstruction}

In this work, we extract multi-view features from each CT image and expect that they provide complementary information for the severity assessment of COVID-19. To flexibly integrate them into a latent space and effectively exploit the heterogeneous information, we propose a multi-view representation learning framework here.

As for the input data in our method, each instance is a multi-view sample, i.e., $\mathbf{x}_i = [ \mathbf{x}_i^{(1)}, \cdots, \mathbf{x}_i^{(V)} ]$, where $V$ is the number of feature types ($V=7$ in our experiments). From the reconstruction perspective, a multi-view representation $\mathbf{z}_i$ is complete if any view $\mathbf{x}_i^{(v)}$ in an observation $\mathbf{x}_i$ can be well reconstructed by a mapping $R_v(\cdot)$, i.e., $\mathbf{x}_i^{(v)} = R_v(\mathbf{z}_i)$ \cite{zhang2019cpm}.

To guarantee the completeness of information, we adopt a multi-view reconstruction loss as follows. 
It consists of two terms. The first one is placed on the source domain as
\begin{equation}
\label{equ:reconstruction loss on the source domain}
\mathcal{L}_r^s = \frac{1}{n_s} \sum\nolimits_{v=1}^{V} \left \| R_v (\mathbf{Z}_{s}) - \mathbf{X}_s'^{(v)} \right \|_F^2,
\end{equation}
where $\left \| \cdot \right \|_F$ is the Frobenius norm, $\mathbf{Z}_{s} = F \circ D (\mathbf{X}_{s}) = F (\mathbf{X}_s')$, $\mathbf{X}_s'^{(v)}$ is $v$-th type of feature in $\mathbf{X}_s'$, and the multi-view reconstruction networks $\left \{ R_v(\cdot) \right \}_{i=1}^{V}$ are utilized in parallel as shown in Fig. \ref{fig:Overall architecture of the FLARE}. The second term is a reconstruction loss on the target domain as
\begin{equation}
\label{equ:reconstruction loss on the target domain}
\begin{aligned}
\mathcal{L}_r^t = & \frac{1}{n_t} \sum\nolimits_{v=1}^{V} \left \| R_v (\mathbf{Z}_{t}) - \mathbf{X}_{t}^{(v)} \right \|_F^2 \\
                   &+ \frac{1}{n_u} \sum\nolimits_{v=1}^{V} \left \| R_v (\mathbf{Z}_{u}) - \mathbf{X}_{u}^{(v)} \right \|_F^2,\\
\end{aligned}
\end{equation}
where $\mathbf{Z}_{t} = F (\mathbf{X}_{t})$, $\mathbf{Z}_{u} = F (\mathbf{X}_{u})$, $\mathbf{X}_t^{(v)}$ and $\mathbf{X}_u^{(v)}$ are the $v$-th type of features in $\mathbf{X}_t$ and $\mathbf{X}_u$, respectively.

The advantages of multi-view learning are two-fold. Firstly, such a reconstruction ensures that the feature extractor $F (\cdot)$ is injective or approximately injective as the autoencoder module designed in \cite{ren2019learning}. Secondly, it ensures completeness of the latent space, resulting in full use of the extracted heterogeneous features.

\subsection{Total Objective Function}
\label{subsec:Total Objective Function}

The final loss in the FLARE method is
\begin{equation}
\label{equ:final loss}
\mathcal{L}_f = \lambda_1 (\mathcal{L}_c^s + \mathcal{L}_c^t) + \lambda_2 \mathcal{L}_p + \lambda_3 (\mathcal{L}_r^s + \mathcal{L}_r^t),
\end{equation}
where $\lambda_1$, $\lambda_2$, and $\lambda_3$ are the trade-off parameters. 
FLARE learns a desirable space with domain-transferability, discriminant, and completeness in an end-to-end manner. 
The backpropagation algorithm is used to optimize the loss function \eqref{equ:final loss}, which can be implemented effectively by TensorFlow or PyTorch, etc. Therefore, we omit the derivation details here.

\subsection{Extension to Multi-source Domain Adaptation}
\label{subsec:extension to multi-source DA}

{In practical scenarios, it is likely that we have labeled data from multiple sites. Owing to the complementary knowledge among these sites, transferring discriminant information from multiple source domains to a target domain is expected to improve the performance of single-source models. However, the multiple source domains contain possible domain shifts with each other. Directly applying single-source DA methods on the combination of all source domain data may lead to negative transfer. To tackle this challenge, we extend the FLARE method to the multi-source adaptation scenario and denote it as M-FLARE for abbreviation.}

Suppose $\{ \mathcal{D}_{s,e}\}_{e=1}^E$ are multiple distinct source domains, where $E$ is the number of source domains. Domain shifts may exist not only between $\mathcal{D}_{s,e}$ and $\mathcal{D}_t$, but also between the source domains. Hence, we assign each source domain $\mathcal{D}_{s,e}$ a different domain translator $D_e(\cdot)$ and a classifier $L_e(\cdot)$, whereas the feature extractor $F(\cdot)$ is shared across all domains to learn domain-invariant features as \cite{xu2018deep, peng2019moment, kang2020contrastive}. For the target domain, data $\mathbf{X}_{u}$ is forwarded into $F(\cdot)$ to extract feature $\mathbf{Z}_{u}$, which is then fed into the multiple classifiers $\{L_e(\cdot)\}$. Finally, we combine the $E$ predictions by a weighted rule:
\begin{equation}
\label{eq:target prediction on multiple source classifiers}
\hat{Y}_u = \sum\nolimits_{e=1}^{E} w_e L_e(\mathbf{Z}_{u}),
\end{equation}
where $w_e$ is the weight of the $e$-th source domain and is defined hereinafter.

On the basis of the above framework, the overall objective function of M-FLARE is defined as: 
\begin{equation}
\label{equ:final multi_source DA loss}
\mathcal{L}_f^m = \lambda_1 \mathcal{L}_c^t + \lambda_3 \mathcal{L}_r^t + \sum_{e=1}^E (\frac{\lambda_1}{E} \mathcal{L}_c^{s,e} + \lambda_2 w_e \mathcal{L}_p^{e} + \frac{\lambda_3}{E} \mathcal{L}_r^{s,e}),
\end{equation}
where $\lambda_1$, $\lambda_2$, and $\lambda_3$ are the trade-off parameters; $\mathcal{L}_c^t$ and $\mathcal{L}_r^t$ are calculated by \eqref{equ:CMMD loss on the target domain} and \eqref{equ:reconstruction loss on the target domain}, respectively; $\mathcal{L}_c^{s,e}$, $\mathcal{L}_p^{e}$, and $\mathcal{L}_r^{s,e}$ are calculated by \eqref{equ:CMMD loss on the source domain}, \eqref{equ:prototype triplet loss}, and \eqref{equ:reconstruction loss on the source domain}, respectively, for the $e$-th source domain. At the training phase, the value of the CMMD loss $\mathcal{L}_c^{s,e}$ can represent how well the model fits into data in the $e$-th source domain. Further, when some source domain is fitted well, we expect to assign a larger weight for the alignment between this source domain and the target domain. With the above analysis, we define $w_e = (\mathcal{L}_c^{s,e})^{-1} / \sum_{j=1}^E (\mathcal{L}_c^{s,j})^{-1}$ and place it in \eqref{eq:target prediction on multiple source classifiers} and \eqref{equ:final multi_source DA loss}. In our experiments, we initialize $w_e$ using $1/E$ and update it at the $(t+1)$-th epoch based on the values of $\mathcal{L}_c^{s,e}$ at the $t$-th epoch.

\section{Experiments}
\label{sec:Experiments}

\subsection{Datasets}
\label{subsec:Datasets}

We perform extensive experiments on both private and public datasets. In terms of private datasets, 5,358 CT images are collected from three clinical sites, i.e., the Second Xiangya Hospital of Central South University, Hangzhou First People's Hospital of Zhejiang University, and Southwest Hospital of Third Military Medical University (Army Medical University). We denote these sites as Site-1, Site-2, and Site-3, respectively. As \cite{tang2021severity}, the clinical status of patients is divided into two groups: mild group (mild or common case) and severe group (severe, critical, or dead case). In addition, we evaluate our method on a publicly available dataset MosMedData~\cite{morozov2020mosmeddata}, which is collected by Moscow City Health Care and denoted as Site-4. The original MosMedData contains 1,110 CT scans with five categories. To perform the same classification task, we divide 856 symptomatic cases into two groups according to the lung lesion grading and routing rules in~\cite{morozov2020mosmeddata}: mild group (mild or moderate case) and severe group (severe or critical case). Following the previous protocols \cite{kang2020diagnosis, shi2021large}, we extract 237-dimensional multi-view heterogeneous features from each CT image which contains seven groups: gray features, texture features, histogram features, number features, intensity features, surface features, and volume features. A standardization preprocessing is adopted before using these multi-view features.

Since the clinical status groups (mild and severe groups) are usually imbalanced, directly evaluating a model on the data may lead to the failure of metrics, e.g., a misleading high accuracy score. Here, {we conduct experiments under both balanced and imbalanced settings.} As regards the balanced setting, we first randomly divide the severe group into 30\% and 70\% for the training and testing set. Then, we randomly select the same number of mild group samples for forming a balanced testing set. The rest of the mild group samples are used as training data. As regards the imbalanced setting, we randomly divide all data into 30\% and 70\% for the training and testing set. We summarize the datasets in Table \ref{tab:Detailed information of datasets.}.

We evaluate FLARE on three transfer tasks between the private datasets: (S1,S2)$\rightarrow$S1, (S2,S1)$\rightarrow$S2, and (S3,S1)$\rightarrow$S3. Let us take the task (S1,S2)$\rightarrow$S1 as an example. The three domains are: a labeled target domain containing the training set of Site-1, a source domain comprised of all labeled data in Site-2, and an unlabeled target domain consisting of the testing set of Site-1. In addition, we implement transfer task from private dataset to public dataset: (S4,S1)$\rightarrow$S4. Note that a different dataset partition setting may result in different classification performance. We adopt random partition ten times and report the average evaluation metrics and the corresponding standard errors.

\begin{table}[!t]
\caption{Detailed information of datasets. Site-1, Site-2, and Site-3 are private. Site-4 is publicly available~\cite{morozov2020mosmeddata}.}
\label{tab:Detailed information of datasets.}
\begin{center}
\begin{tabular}{c||cc|cc}
\hline
 & \multicolumn{2}{c|}{Training set} & \multicolumn{2}{c}{Testing set} \\
 & Mild group & Severe group & Mild group & Severe group \\
\hline
\multicolumn{5}{c}{Balanced setting} \\
\hline
Site-1 & 2230 & 88 & 203 & 203 \\
Site-2 & 1055 & 34 & 78 & 78 \\
Site-3 & 317 & 190 & 441 & 441 \\
Site-4 & 777 & 15 & 32 & 32 \\
\hline
\hline
\multicolumn{5}{c}{Imbalanced setting} \\
\hline
Site-1 & 730 & 88 & 1703 & 203 \\
Site-2 & 340 & 34 & 793 & 78 \\
Site-3 & 228 & 190 & 530 & 441 \\
Site-4 & 243 & 15 & 566 & 32 \\
\hline
\end{tabular}
\end{center}
\end{table}

\subsection{Experimental Setup}
\label{subsec:Experimental Setup}

\subsubsection{Implementation details}
With regard to the domain translator, feature extractor, classifier, and reconstruction networks, we use FCNs with 2, 1, 1, and 1 hidden layers, respectively. All the experiments are implemented in PyTorch 1.0.1. AMSGrad \cite{reddi2018convergence} optimizer with learning rate $lr = 2e{\text -}4$, $\beta_{1} = 0.5$, $\beta_{2} = 0.999$, and batch size of 100 is applied over 300 epochs. Since the data flow is different between source and target domains (the domain translator is employed only for the former), the optimization process is implemented alternately on the two domains. Specifically, in each epoch, we randomly divide both source and target data into several batches. Based on each mini-batch, we optimize the objective function on the target domain and source domain in turn. Moreover, we fix $\lambda_1 = 1$ and tune the other three hyper-parameters using grid search: $\lambda_2 \in \{1e1, 5e1, 1e2, 5e2, 1e3\}$, $\lambda_3 \in \{2e{\text -}4, 1e{\text -}3, 2e{\text -}3, 5e{\text -}3, 2e{\text -}2\}$, $\alpha \in \{0, 0.4, 0.8, 1.2, 1.6\}$. We select the hyper-parameters based on 5-fold cross-validation on the target labeled data.

\begin{table*}
\caption{Comparisons against competing methods on DA tasks between private datasets. Left to right: three DA tasks (S1,S2)$\rightarrow$S1, (S2,S1)$\rightarrow$S2, and (S3,S1)$\rightarrow$S3. Our FLARE outperforms all the three groups of methods in the overall performance metrics, i.e., ${\rm F1}$, and ${\rm G}{\text -}{\rm mean}$.}
\label{tab:Comparisons_SOTA_1}
\renewcommand\tabcolsep{3.6pt}
\scriptsize
\begin{center}
\begin{tabular}{c|c|| llll| llll| llll}
\hline
\multicolumn{2}{c||}{ } & \multicolumn{4}{c|}{(S1,S2)$\rightarrow$S1} & \multicolumn{4}{c|}{(S2,S1)$\rightarrow$S2} & \multicolumn{4}{c}{(S3,S1)$\rightarrow$S3} \\
\multicolumn{2}{c||}{Methods} & \multicolumn{1}{c}{${\rm SEN}$} & \multicolumn{1}{c}{${\rm SPE}$} & \multicolumn{1}{c}{${\rm F1}$} & \multicolumn{1}{c|}{${\rm G}{\text -}{\rm mean}$} & \multicolumn{1}{c}{${\rm SEN}$} & \multicolumn{1}{c}{${\rm SPE}$} & \multicolumn{1}{c}{${\rm F1}$} & \multicolumn{1}{c|}{${\rm G}{\text -}{\rm mean}$} & \multicolumn{1}{c}{${\rm SEN}$} & \multicolumn{1}{c}{${\rm SPE}$} & \multicolumn{1}{c}{${\rm F1}$} & \multicolumn{1}{c}{${\rm G}{\text -}{\rm mean}$} \\
\hline
\multicolumn{14}{c}{Balanced setting} \\
\hline
\multirow{10}{0.5cm}[0cm]{\rotatebox{90}{\parbox{2.4cm}{Target-only supervised \\learning methods}}}
& LR & 24.6$\pm$0.8 & \textbf{99.9$\pm$0.1} & 39.4$\pm$1.1 & 49.5$\pm$0.9 & 38.0$\pm$1.1 & 99.9$\pm$0.1 & 54.9$\pm$1.1 & 61.5$\pm$0.9 & 54.3$\pm$2.5 & 80.6$\pm$2.2 & 62.2$\pm$1.2 & 65.8$\pm$0.7  \\
& SVM & 37.1$\pm$2.3 & 98.4$\pm$0.4 & 53.1$\pm$2.4 & 60.1$\pm$1.9 & 41.5$\pm$2.0 & 99.9$\pm$0.1 & 58.4$\pm$1.9 & 64.3$\pm$1.5 & 48.1$\pm$2.5 & 78.4$\pm$2.5 & 56.4$\pm$1.5 & 61.0$\pm$0.9  \\
& GNB & 39.8$\pm$11.0 & 90.3$\pm$2.9 & 72.4$\pm$1.7 & 44.6$\pm$12.1 & 59.5$\pm$10.1 & 90.6$\pm$1.8 & 79.8$\pm$1.4 & 64.7$\pm$10.8 & {\tiny~~}0.1$\pm$0.0 & {\tiny~}\textbf{100$\pm$0.0} & {\tiny~~}0.5$\pm$0.0 & {\tiny~~}1.4$\pm$0.8  \\
& KNN & {\tiny~~}9.0$\pm$0.6 & 99.9$\pm$0.1 & 16.5$\pm$1.0 & 29.9$\pm$1.0 & 10.8$\pm$1.4 & {\tiny~}\textbf{100$\pm$0.0} & 19.2$\pm$2.2 & 32.2$\pm$2.2 & 56.9$\pm$1.3 & 77.6$\pm$1.1 & 63.4$\pm$0.8 & 66.3$\pm$0.6  \\
& FCN & 24.9$\pm$6.4 & 83.7$\pm$4.2 & 31.9$\pm$6.0 & 41.7$\pm$4.6 & 29.2$\pm$7.6 & 69.1$\pm$7.6 & 32.9$\pm$6.7 & 39.8$\pm$5.2 & 63.4$\pm$0.7 & 74.8$\pm$1.2 & 67.2$\pm$0.4 & 68.8$\pm$0.4  \\
& LR+RanOver & 51.1$\pm$1.6 & 96.8$\pm$0.7 & 66.1$\pm$1.2 & 70.2$\pm$0.9 & 65.4$\pm$2.1 & 99.6$\pm$0.3 & 78.7$\pm$1.5 & 80.6$\pm$1.3 & 57.9$\pm$2.2 & 76.1$\pm$2.4 & 63.5$\pm$1.0 & 66.0$\pm$0.6  \\
& SVM+RanOver & 39.1$\pm$2.5 & 97.6$\pm$0.6 & 54.8$\pm$2.4 & 61.4$\pm$1.8 & 41.5$\pm$2.0 & 99.9$\pm$0.1 & 58.4$\pm$1.9 & 64.3$\pm$1.5 & 48.8$\pm$2.7 & 77.6$\pm$3.0 & 56.7$\pm$1.6 & 61.0$\pm$1.1  \\
& GNB+RanOver & 40.6$\pm$11.1 & 89.5$\pm$3.0 & 62.6$\pm$10.2 & 45.6$\pm$11.8 & 60.3$\pm$10.2 & 89.6$\pm$2.0 & 79.9$\pm$1.3 & 64.7$\pm$10.8 & {\tiny~~}0.3$\pm$0.1 & {\tiny~}\textbf{100$\pm$0.0} & {\tiny~~}0.9$\pm$0.2 & {\tiny~~}3.9$\pm$1.2  \\
& KNN+RanOver & 52.4$\pm$1.3 & 97.7$\pm$0.2 & 67.7$\pm$1.1 & 71.5$\pm$0.9 & 57.1$\pm$2.2 & 99.5$\pm$0.3 & 72.2$\pm$1.8 & 75.2$\pm$1.5 & 68.4$\pm$1.0 & 65.8$\pm$1.0 & 67.5$\pm$0.5 & 67.0$\pm$0.4  \\
& FCN+SCBS & 72.7$\pm$1.4 & 88.5$\pm$1.3 & 78.9$\pm$0.6 & 80.1$\pm$0.5 & 79.7$\pm$1.8 & 93.3$\pm$1.9 & 85.5$\pm$0.7 & 86.1$\pm$0.6 & 69.1$\pm$1.6 & 69.7$\pm$2.4 & 69.2$\pm$0.6 & 69.1$\pm$0.7  \\
\hline          
\multirow{10}{0.5cm}[0.83cm]{\rotatebox{90}{\parbox{1.0cm}{Unsuper. \\DA meth.}}}
& Source-only & 53.6$\pm$1.7 & 82.0$\pm$0.9 & 62.3$\pm$1.3 & 66.2$\pm$1.1 & 63.2$\pm$1.7 & 86.9$\pm$1.1 & 71.6$\pm$1.1 & 74.0$\pm$0.9 & 43.2$\pm$1.0 & 83.8$\pm$0.5 & 54.1$\pm$0.8 & 60.1$\pm$0.5 \\
& DAN \cite{long2015learning} & \textbf{80.3$\pm$1.0} & 66.3$\pm$1.1 & 75.0$\pm$0.5 & 72.9$\pm$0.5 & 81.2$\pm$1.5 & 80.6$\pm$1.8 & 80.9$\pm$1.1 & 80.8$\pm$1.1 & 64.6$\pm$0.8 & 67.3$\pm$1.0 & 65.5$\pm$0.6 & 65.9$\pm$0.6 \\
& DANN \cite{ganin2016domain} & 76.7$\pm$1.2 & 70.4$\pm$1.1 & 74.3$\pm$0.5 & 73.4$\pm$0.4 & 78.2$\pm$1.4 & 82.1$\pm$2.0 & 79.7$\pm$0.8 & 80.0$\pm$0.9 & 61.7$\pm$1.1 & 71.7$\pm$1.4 & 64.9$\pm$0.6 & 66.4$\pm$0.5 \\
& CDAN+E \cite{long2018conditional} & 77.3$\pm$1.1 & 70.2$\pm$1.3 & 74.7$\pm$0.3 & 73.6$\pm$0.3 & 83.1$\pm$1.4 & 80.1$\pm$1.5 & 81.9$\pm$0.9 & 81.5$\pm$0.9 & 61.0$\pm$1.0 & 71.8$\pm$1.0 & 64.5$\pm$0.6 & 66.1$\pm$0.5 \\
\hline          
\multirow{10}{0.5cm}[0.23cm]{\rotatebox{90}{\parbox{1.7cm}{Semi-supervised \\DA methods}}}
& JointDomain & 74.6$\pm$0.9 & 80.4$\pm$1.1 & 76.8$\pm$0.4 & 77.4$\pm$0.4 & 78.3$\pm$1.3 & 91.4$\pm$0.9 & 83.8$\pm$0.9 & 84.6$\pm$0.8 & 62.9$\pm$1.0 & 76.2$\pm$0.9 & 67.3$\pm$0.6 & 69.2$\pm$0.5  \\
& DACoM \cite{li2018semi} & 74.9$\pm$1.9 & 73.2$\pm$1.8 & 74.2$\pm$0.9 & 73.9$\pm$0.8 & 80.9$\pm$1.7 & 83.5$\pm$1.3 & 81.9$\pm$1.0 & 82.1$\pm$0.9 & 65.6$\pm$1.5 & 67.6$\pm$1.8 & 66.2$\pm$0.7 & 66.4$\pm$0.7  \\
& CDFSB \cite{chen2019a} & 74.3$\pm$1.7 & 83.6$\pm$1.4 & 77.9$\pm$0.6 & 78.7$\pm$0.4 & 74.6$\pm$2.2 & 93.5$\pm$1.5 & 82.3$\pm$1.2 & 83.4$\pm$1.0 & 69.8$\pm$1.3 & 62.5$\pm$1.8 & 67.3$\pm$0.5 & 65.9$\pm$0.6  \\
& CDFSB++ \cite{chen2019a} & 79.4$\pm$1.0 & 74.5$\pm$1.0 & 77.5$\pm$0.5 & 76.9$\pm$0.5 & 81.0$\pm$0.7 & 86.4$\pm$0.9 & 83.3$\pm$0.5 & 83.7$\pm$0.5 & 70.5$\pm$1.4 & 63.6$\pm$1.4 & 68.1$\pm$0.7 & 66.8$\pm$0.6  \\
& MME \cite{saito2019semi} & 74.6$\pm$0.6 & 90.4$\pm$1.2 & 81.0$\pm$0.5 & 82.1$\pm$0.5 & 82.4$\pm$1.2 & 93.5$\pm$0.9 & 87.2$\pm$0.6 & 87.7$\pm$0.5 & 60.9$\pm$0.9 & 75.3$\pm$1.4 & 65.6$\pm$0.4 & 67.6$\pm$0.4  \\
& LFT \cite{Tseng2020Cross-Domain} & 73.2$\pm$1.8 & 90.5$\pm$2.2 & 80.1$\pm$0.8 & 81.2$\pm$0.7 & 81.7$\pm$2.4 & 93.8$\pm$2.1 & 86.9$\pm$0.9 & 87.3$\pm$0.8 & 68.3$\pm$0.9 & 71.7$\pm$1.3 & 69.5$\pm$0.5 & 69.9$\pm$0.5  \\
& CCSL \cite{wang2020contrastive} & 73.6$\pm$1.0 & 88.3$\pm$1.4 & 79.4$\pm$0.4 & 80.5$\pm$0.4 & 81.0$\pm$1.3 & 96.0$\pm$1.4 & 87.6$\pm$0.9 & 88.2$\pm$0.8 & 64.6$\pm$1.4 & 72.6$\pm$1.6 & 67.2$\pm$0.7 & 68.3$\pm$0.5  \\
& \cellcolor{gray!20}\textbf{FLARE} & \cellcolor{gray!20}79.7$\pm$0.8 & \cellcolor{gray!20}94.1$\pm$0.8 & \cellcolor{gray!20}\textbf{85.8$\pm$0.4} & \cellcolor{gray!20}\textbf{86.5$\pm$0.4} & \cellcolor{gray!20}\textbf{88.1$\pm$1.2} & \cellcolor{gray!20}94.6$\pm$0.7 & \cellcolor{gray!20}\textbf{91.0$\pm$0.7} & \cellcolor{gray!20}\textbf{91.3$\pm$0.7} & \cellcolor{gray!20}\textbf{71.4$\pm$1.5} & \cellcolor{gray!20}71.5$\pm$2.1 & \cellcolor{gray!20}\textbf{71.4$\pm$0.3} & \cellcolor{gray!20}\textbf{71.2$\pm$0.4}  \\
\hline  
\hline         
\multicolumn{14}{c}{Imbalanced setting} \\
\hline      
\multirow{10}{0.5cm}[0cm]{\rotatebox{90}{\parbox{2.4cm}{Target-only supervised \\learning methods}}}
& LR & 60.5$\pm$5.5 & \textbf{94.6$\pm$1.2} & 59.1$\pm$0.2 & 75.5$\pm$3.0 & 39.2$\pm$3.9 & 99.7$\pm$0.1 & 54.2$\pm$4.0 & 61.8$\pm$3.1 & 69.2$\pm$0.5 & 63.6$\pm$1.5 & 65.1$\pm$0.5 & 66.4$\pm$0.7  \\
& SVM & 62.2$\pm$3.8 & 90.4$\pm$1.2 & 51.0$\pm$0.9 & 74.5$\pm$1.9 & 41.4$\pm$3.7 & 99.6$\pm$0.1 & 56.1$\pm$3.8 & 63.6$\pm$2.9 & 69.7$\pm$2.2 & 53.6$\pm$2.2 & 61.8$\pm$0.8 & 60.8$\pm$0.7  \\
& GNB & 51.4$\pm$12.3 & 73.2$\pm$12.4 & 38.3$\pm$5.0 & 36.7$\pm$12.3 & 37.9$\pm$7.3 & 95.2$\pm$1.1 & 45.0$\pm$2.1 & 52.8$\pm$9.1 & 15.1$\pm$9.9 & 91.5$\pm$5.7 & 19.7$\pm$11.9 & 16.7$\pm$8.2  \\
& KNN & 19.3$\pm$1.5 & 99.5$\pm$0.1 & 30.9$\pm$1.8 & 43.5$\pm$1.7 & 5.6$\pm$1.5 & \textbf{99.9$\pm$0.0} & 12.6$\pm$2.6 & 20.0$\pm$4.2 & 65.0$\pm$2.1 & 68.5$\pm$1.4 & 63.9$\pm$0.9 & 66.5$\pm$0.5  \\
& FCN & 44.8$\pm$5.7 & 72.5$\pm$5.3 & 24.3$\pm$1.0 & 54.2$\pm$2.6 & 45.5$\pm$6.8 & 74.9$\pm$4.8 & 23.1$\pm$2.7 & 55.5$\pm$4.4 & 70.2$\pm$0.8 & 62.1$\pm$0.9 & 65.2$\pm$0.4 & 66.0$\pm$0.3  \\
& LR+RanOver & 60.4$\pm$2.2 & 93.5$\pm$0.5 & 56.0$\pm$1.4 & 75.0$\pm$1.3 & 48.0$\pm$1.8 & 99.0$\pm$0.4 & 60.5$\pm$1.5 & 68.9$\pm$1.2 & 70.2$\pm$0.7 & 62.3$\pm$2.8 & 65.3$\pm$0.7 & 66.1$\pm$1.2  \\
& SVM+RanOver & 62.4$\pm$3.9 & 90.1$\pm$1.2 & 50.6$\pm$1.1 & 74.5$\pm$2.0 & 41.4$\pm$3.7 & 99.6$\pm$0.1 & 56.1$\pm$3.8 & 63.6$\pm$2.9 & 71.2$\pm$2.2 & 51.9$\pm$2.4 & 62.1$\pm$0.7 & 60.4$\pm$0.8  \\
& GNB+RanOver & 51.2$\pm$12.3 & 73.2$\pm$12.4 & 38.2$\pm$5.0 & 36.6$\pm$12.2 & 38.7$\pm$7.4 & 94.8$\pm$1.2 & 44.8$\pm$1.9 & 53.3$\pm$9.2 & 15.2$\pm$9.8 & \textbf{91.6$\pm$5.6} & 17.6$\pm$10.5 & 18.2$\pm$8.0  \\
& KNN+RanOver & 63.3$\pm$1.7 & 90.1$\pm$1.1 & 51.6$\pm$1.5 & 75.4$\pm$0.8 & 38.9$\pm$3.0 & 96.8$\pm$0.3 & 44.8$\pm$2.3 & 60.9$\pm$2.4 & 69.2$\pm$1.8 & 63.3$\pm$1.4 & 64.8$\pm$0.8 & 66.0$\pm$0.6  \\
& FCN+SCBS & 69.0$\pm$1.6 & 92.0$\pm$0.8 & 58.6$\pm$1.6 & 79.6$\pm$0.9 & 74.5$\pm$2.6 & 89.4$\pm$1.5 & 54.1$\pm$2.4 & 81.4$\pm$1.3 & 71.8$\pm$1.1 & 62.2$\pm$2.2 & 66.0$\pm$0.2 & 66.8$\pm$0.7  \\
\hline          
\multirow{10}{0.5cm}[0.83cm]{\rotatebox{90}{\parbox{1.0cm}{Unsuper. \\DA meth.}}}
& Source-only & 55.4$\pm$1.5 & 80.4$\pm$0.6 & 34.4$\pm$0.5 & 66.6$\pm$0.7 & 69.2$\pm$2.2 & 86.5$\pm$1.1 & 45.7$\pm$1.5 & 77.2$\pm$1.0 & 41.2$\pm$0.9 & 84.6$\pm$0.5 & 51.6$\pm$0.8 & 59.0$\pm$0.6  \\
& DAN \cite{long2015learning} & 79.9$\pm$0.6 & 68.0$\pm$0.6 & 35.4$\pm$0.6 & 73.7$\pm$0.3 & 84.2$\pm$1.4 & 78.0$\pm$0.7 & 41.7$\pm$1.0 & 81.0$\pm$0.6 & 62.4$\pm$1.2 & 69.3$\pm$1.0 & 62.6$\pm$0.7 & 65.7$\pm$0.5  \\
& DANN \cite{ganin2016domain} & 79.8$\pm$0.5 & 67.2$\pm$0.7 & 34.9$\pm$0.5 & 73.3$\pm$0.3 & 81.2$\pm$1.1 & 82.6$\pm$1.3 & 46.1$\pm$1.8 & 81.9$\pm$0.7 & 62.4$\pm$0.9 & 71.2$\pm$1.0 & 63.3$\pm$0.5 & 66.6$\pm$0.4  \\
& CDAN+E \cite{long2018conditional} & 80.0$\pm$0.6 & 68.2$\pm$0.6 & 35.6$\pm$0.5 & 73.9$\pm$0.3 & 79.0$\pm$2.0 & 81.8$\pm$0.8 & 43.8$\pm$0.7 & 80.3$\pm$0.9 & 59.4$\pm$0.8 & 73.4$\pm$1.0 & 62.1$\pm$0.4 & 66.0$\pm$0.3  \\
\hline          
\multirow{10}{0.5cm}[0.23cm]{\rotatebox{90}{\parbox{1.7cm}{Semi-supervised \\DA methods}}}
& JointDomain & 78.0$\pm$1.2 & 78.5$\pm$0.5 & 43.3$\pm$0.4 & 78.2$\pm$0.5 & 70.2$\pm$2.4 & 91.5$\pm$0.9 & 55.2$\pm$1.8 & 80.0$\pm$1.2 & 64.8$\pm$0.8 & 70.0$\pm$0.8 & 64.6$\pm$0.4 & 67.4$\pm$0.3  \\
& DACoM \cite{li2018semi} & 71.6$\pm$1.6 & 77.3$\pm$1.8 & 39.8$\pm$1.3 & 74.3$\pm$0.7 & 80.3$\pm$2.0 & 82.8$\pm$2.0 & 46.8$\pm$2.4 & 81.4$\pm$1.1 & 66.3$\pm$1.8 & 67.1$\pm$2.0 & 64.4$\pm$0.5 & 66.5$\pm$0.3  \\
& CDFSB \cite{chen2019a} & 77.0$\pm$1.2 & 80.9$\pm$0.8 & 45.6$\pm$0.7 & 78.9$\pm$0.5 & 75.4$\pm$1.1 & 88.2$\pm$1.1 & 51.9$\pm$1.8 & 81.5$\pm$0.6 & 69.8$\pm$1.5 & 59.1$\pm$1.5 & 63.7$\pm$0.6 & 64.0$\pm$0.4  \\
& CDFSB++ \cite{chen2019a} & 79.6$\pm$1.0 & 71.7$\pm$1.0 & 38.7$\pm$0.6 & 75.5$\pm$0.2 & 81.5$\pm$1.5 & 86.5$\pm$0.7 & 51.6$\pm$0.8 & 83.9$\pm$0.6 & 69.9$\pm$1.3 & 64.7$\pm$1.1 & 65.8$\pm$0.6 & 67.2$\pm$0.5  \\
& MME \cite{saito2019semi} & 72.4$\pm$1.3 & 88.4$\pm$1.2 & 54.0$\pm$1.6 & 79.9$\pm$0.4 & 80.7$\pm$1.8 & 92.6$\pm$0.6 & 63.5$\pm$1.8 & 86.4$\pm$1.0 & 59.5$\pm$3.1 & 74.7$\pm$2.7 & 62.7$\pm$1.4 & 66.5$\pm$0.7  \\
& LFT \cite{Tseng2020Cross-Domain} & 73.6$\pm$1.5 & 89.2$\pm$2.2 & 57.7$\pm$3.2 & 80.8$\pm$0.8 & 80.9$\pm$1.3 & 91.3$\pm$1.8 & 63.0$\pm$4.2 & 85.9$\pm$1.0 & 69.1$\pm$1.0 & 66.6$\pm$1.3 & 66.1$\pm$0.7 & 67.8$\pm$0.6  \\
& CCSL \cite{wang2020contrastive} & 76.2$\pm$1.0 & 85.6$\pm$1.1 & 51.5$\pm$1.4 & 80.7$\pm$0.5 & 80.2$\pm$1.6 & 94.6$\pm$0.7 & 68.7$\pm$1.5 & 87.0$\pm$0.8 & 65.9$\pm$1.1 & 70.5$\pm$1.0 & 65.4$\pm$0.6 & 68.1$\pm$0.4  \\
& \cellcolor{gray!20}\textbf{FLARE} & \cellcolor{gray!20}\textbf{80.7$\pm$1.3} & \cellcolor{gray!20}92.3$\pm$0.8 & \cellcolor{gray!20}\textbf{66.1$\pm$1.5} & \cellcolor{gray!20}\textbf{86.3$\pm$0.6} & \cellcolor{gray!20}\textbf{87.1$\pm$1.7} & \cellcolor{gray!20}95.2$\pm$1.5 & \cellcolor{gray!20}\textbf{76.0$\pm$3.4} & \cellcolor{gray!20}\textbf{91.0$\pm$0.8} & \cellcolor{gray!20}\textbf{72.5$\pm$1.2} & \cellcolor{gray!20}68.0$\pm$1.2 & \cellcolor{gray!20}\textbf{68.7$\pm$0.5} & \cellcolor{gray!20}\textbf{70.1$\pm$0.4}  \\
\hline
\end{tabular}
\end{center}
\end{table*}

\subsubsection{Evaluation metrics}
The severe infection case is denoted as the positive class. The default value of the operating point is set as 0.5. For evaluating the severity assessment performance of COVID-19, four standard metrics including ${\rm SEN}$ (sensitivity), ${\rm SPE}$ (specificity), ${\rm F1}$ (harmonic mean of sensitivity and precision), and ${\rm G}{\text -}{\rm mean}$ (geometric mean of sensitivity and specificity) are used.

\subsection{Comparison With the State-of-the-art Methods}
\label{subsec:Comparison with SOTA}
We present extensive comparisons between FLARE and several other methods under both balanced and imbalanced settings. These methods are divided into three groups. The first group consists of five target-only supervised learning methods, i.e., logistic regression (LR), support vector machine (SVM), Gaussian-naive-Bayes (GNB), k-nearest-neighbors (KNN), and FCN. To relieve the class-imbalance problem, we adopt random oversampling (RanOver for short) and SCBS strategies for them. The second group has an approach directly trained on source data (Source-only for short) and some recent unsupervised DA methods: deep adaptation network (DAN) \cite{long2015learning}, domain-adversarial neural network (DANN) \cite{ganin2016domain}, and conditional domain adversarial network (CDAN) \cite{long2018conditional}. The last group includes an algorithm trained jointly using source and labeled target data (JointDomain for short), and some semi-supervised DA methods, i.e., DACoM \cite{li2018semi}, cross-domain few-shot baseline (CDFSB) method \cite{chen2019a}, CDFSB++ \cite{chen2019a}, MME \cite{saito2019semi}, learned feature-wise transformation (LFT) based classification algorithm \cite{Tseng2020Cross-Domain}, and contrastive cross-site learning (CCSL) method \cite{wang2020contrastive}. For fair comparisons, we adopt the same imbalanced learning strategy and network architecture, and place a multi-view reconstruction loss in the representation learning phase for FCN and all DA methods, since the features used in these experiments include heterogeneous types.

\subsubsection{Transfer tasks between the private datasets}
Table \ref{tab:Comparisons_SOTA_1} reports the results on transfer tasks between the private datasets. 
FLARE yields compelling results on all three DA tasks. On the task (S1,S2)$\rightarrow$S1 with balanced setting, DAN \cite{long2015learning} and LR obtain better sensitivity and specificity measures, respectively, whereas FLARE outperforms all approaches when assessed by the overall performance metrics, ${\rm F1}$ and ${\rm G}{\text -}{\rm mean}$. On the task (S1,S2)$\rightarrow$S1 with imbalanced setting, LR obtains a better specificity measure, whereas FLARE is best for all the other metrics. Interestingly, the FCN equipping our SCBS strategy outperforms all other target-only learning methods and even the unsupervised DA methods. This demonstrates that the cross-site discrepancy is indeed large, and the supervision information in the target domain is important for classification. FLARE takes the supervision information in both domains into full consideration and achieves excellent results.

On the task (S2,S1)$\rightarrow$S2, KNN exhibits better performance by the specificity measure under both balanced and imbalanced settings, but our FLARE is the best for all others performance metrics. Although all semi-supervised DA methods use the supervision information in source and target domains, FLARE outperforms the others in terms of all metrics (except 1.4\% below the specificity of CCSL \cite{wang2020contrastive} under balanced setting). This effectiveness can be attributed to the prototype triplet loss, where the estimated prototypes are robust to the class-imbalance scenario and the source data is encouraged to be around these prototypes for transferring the discriminant information.

Similarly, FLARE outperforms others for sensitivity, ${\rm F1}$, and ${\rm G}{\text -}{\rm mean}$ measures on the task (S3,S1)$\rightarrow$S3. Although GNB and GNB+RanOver obtain better results in terms of specificity, they exhibit extremely poor performance by the other measures because of the impact of class-imbalance problem. Compared with the results in the first two DA tasks (i.e., Site-1 or Site-2 as the target domain), all methods achieve slightly poorer performance in the last task (i.e., Site-3 as the target domain). We attribute this to the difficulty of the task itself. Particularly, over 30\% of severe patients on the Site-3 encounter critical severe diseases or even deaths, whose lung deteriorations vary owing to the presence of coexisting conditions \cite{guan2020clinical, li2020early}. This causes large within-class heterogeneity, as also visualized in Section \ref{subsec:Visualization}.

\begin{table}
\caption{{Comparisons against competing methods on DA tasks from private to public datasets, i.e., task (S4,S1)$\rightarrow$S4.}}
\label{tab:Comparisons_SOTA_from_private_to_public}
\renewcommand\tabcolsep{4.4pt}
\begin{center}
\begin{tabular}{c|c|| llll}
\hline
\multicolumn{2}{c||}{Methods} & \multicolumn{1}{c}{${\rm SEN}$} & \multicolumn{1}{c}{${\rm SPE}$} & \multicolumn{1}{c}{${\rm F1}$} & \multicolumn{1}{c}{${\rm G}{\text -}{\rm mean}$}  \\
\hline         
\multicolumn{6}{c}{Balanced setting} \\
\hline      
\multirow{10}{0.5cm}[0cm]{\rotatebox{90}{\parbox{2.6cm}{Target-only supervised \\learning methods}}}
& LR & {\tiny~~}3.1$\pm$0.7 & {\tiny~}\textbf{100$\pm$0.0} & {\tiny~~}7.5$\pm$0.9 & 15.6$\pm$2.8  \\
& SVM & {\tiny~~}3.8$\pm$1.0 & 98.8$\pm$0.5 & {\tiny~~}9.9$\pm$1.6 & 15.8$\pm$3.7  \\
& GNB & {\tiny~~}2.2$\pm$0.5 & {\tiny~}\textbf{100$\pm$0.0} & {\tiny~~}6.1$\pm$0.0 & 12.4$\pm$2.7 \\
& KNN & {\tiny~~}0.0$\pm$0.0 & {\tiny~}\textbf{100$\pm$0.0} & {\tiny~~}0.0$\pm$0.0 & {\tiny~~}0.0$\pm$0.0  \\
& FCN & 38.1$\pm$4.3 & 76.9$\pm$3.6 & 46.4$\pm$4.0 & 53.0$\pm$3.1  \\
& LR+RanOver & {\tiny~~}5.9$\pm$1.1 & 96.0$\pm$0.9 & 11.8$\pm$1.6 & 22.2$\pm$2.9 \\
& SVM+RanOver & {\tiny~~}3.8$\pm$1.0 & 98.8$\pm$0.5 & {\tiny~~}9.9$\pm$1.6 & 15.8$\pm$3.7  \\
& GNB+RanOver & {\tiny~~}2.2$\pm$0.5 & {\tiny~}\textbf{100$\pm$0.0} & {\tiny~~}6.1$\pm$0.0 & 12.4$\pm$2.7 \\
& KNN+RanOver & {\tiny~~}2.8$\pm$1.1 & 96.6$\pm$1.1 & 10.1$\pm$1.9 & 11.3$\pm$3.9  \\
& FCN+SCBS & 43.4$\pm$4.0 & 92.2$\pm$3.1 & 56.7$\pm$3.7 & 62.6$\pm$2.9 \\
\hline          
\multirow{10}{0.5cm}[0.9cm]{\rotatebox{90}{\parbox{1.1cm}{Unsuper. \\DA meth.}}}
& Source-only & 25.9$\pm$1.2 & 97.2$\pm$0.7 & 40.2$\pm$1.4 & 50.1$\pm$1.1 \\
& DAN \cite{long2015learning} & 63.4$\pm$4.0 & 87.8$\pm$3.3 & 71.7$\pm$1.8 & 73.8$\pm$1.2  \\
& DANN \cite{ganin2016domain} & 57.8$\pm$3.7 & 85.2$\pm$2.3 & 66.8$\pm$1.9 & 70.0$\pm$1.2 \\
& CDAN+E \cite{long2018conditional} & 51.6$\pm$2.7 & 94.5$\pm$2.7 & 65.6$\pm$2.3 & 69.7$\pm$1.9 \\
\hline          
\multirow{10}{0.5cm}[0.23cm]{\rotatebox{90}{\parbox{1.9cm}{Semi-supervised \\DA methods}}}
& JointDomain & 30.6$\pm$3.0 & 99.4$\pm$0.4 & 46.0$\pm$3.5 & 54.6$\pm$2.7  \\
& DACoM \cite{li2018semi} & 61.6$\pm$3.1 & 81.6$\pm$2.5 & 68.1$\pm$2.2 & 70.5$\pm$1.8  \\
& CDFSB \cite{chen2019a} & 32.2$\pm$2.9 & 92.5$\pm$1.6 & 45.4$\pm$2.9 & 53.9$\pm$2.2  \\
& CDFSB++ \cite{chen2019a} & 50.0$\pm$1.1 & 97.2$\pm$1.3 & 65.4$\pm$1.1 & 69.7$\pm$0.9  \\
& MME \cite{saito2019semi} & 54.7$\pm$5.1 & 85.0$\pm$2.7 & 63.2$\pm$3.9 & 67.0$\pm$2.9   \\
& LFT \cite{Tseng2020Cross-Domain} & 58.6$\pm$1.5 & 90.6$\pm$2.2 & 69.8$\pm$1.0 & 72.8$\pm$0.9  \\
& CCSL \cite{wang2020contrastive} & 57.5$\pm$4.2 & 88.1$\pm$3.3 & 67.3$\pm$2.4 & 70.3$\pm$1.7  \\
& \cellcolor{gray!20}\textbf{FLARE} & \cellcolor{gray!20}\textbf{68.8$\pm$2.8} & \cellcolor{gray!20}85.9$\pm$2.7 & \cellcolor{gray!20}\textbf{75.9$\pm$1.3} & \cellcolor{gray!20}\textbf{76.4$\pm$0.9} \\
\hline
\hline
\multicolumn{6}{c}{Imbalanced setting} \\
\hline      
\multirow{10}{0.5cm}[0cm]{\rotatebox{90}{\parbox{2.6cm}{Target-only supervised \\learning methods}}}
& LR & {\tiny~~}4.2$\pm$0.7 & 99.9$\pm$0.1 & {\tiny~~}8.7$\pm$1.0 & 19.1$\pm$2.4  \\
& SVM & {\tiny~~}6.3$\pm$1.7 & 97.3$\pm$1.3 & {\tiny~~}8.3$\pm$0.9 & 22.1$\pm$3.4  \\
& GNB & 20.6$\pm$13.2 & 80.0$\pm$13.3 & {\tiny~~}7.8$\pm$1.3 & {\tiny~~}4.2$\pm$2.2  \\
& KNN & {\tiny~~}1.9$\pm$0.7 & {\tiny~}\textbf{100$\pm$0.0} & {\tiny~~}7.2$\pm$1.4 & {\tiny~~}9.6$\pm$3.3  \\
& FCN & 37.0$\pm$6.3 & 70.5$\pm$5.0 & 12.7$\pm$1.7 & 48.3$\pm$2.9  \\
& LR+RanOver & {\tiny~~}8.0$\pm$2.6 & 96.7$\pm$1.7 & {\tiny~~}8.7$\pm$1.2 & 25.3$\pm$3.3  \\
& SVM+RanOver & {\tiny~~}6.3$\pm$1.7 & 97.3$\pm$1.3 & {\tiny~~}8.3$\pm$0.9 & 22.1$\pm$3.4  \\
& GNB+RanOver & 20.6$\pm$13.2 & 80.1$\pm$13.2 & {\tiny~~}7.8$\pm$1.4 & {\tiny~~}5.1$\pm$2.3  \\
& KNN+RanOver & 21.5$\pm$2.2 & 92.9$\pm$1.4 & 18.7$\pm$2.4 & 44.0$\pm$2.3  \\
& FCN+SCBS & 46.2$\pm$5.3 & 92.5$\pm$2.4 & 36.8$\pm$2.8 & 64.0$\pm$3.1  \\
\hline          
\multirow{10}{0.5cm}[0.9cm]{\rotatebox{90}{\parbox{1.1cm}{Unsuper. \\DA meth.}}}
& Source-only & 23.9$\pm$1.8 & 96.5$\pm$0.6 & 26.4$\pm$1.3 & 47.7$\pm$1.7  \\
& DAN \cite{long2015learning} & 61.6$\pm$2.2 & 82.0$\pm$3.5 & 29.3$\pm$3.1 & 71.0$\pm$1.0  \\
& DANN \cite{ganin2016domain} & 55.7$\pm$4.3 & 88.7$\pm$4.1 & 35.8$\pm$5.8 & 70.0$\pm$1.2  \\
& CDAN+E \cite{long2018conditional} & 57.0$\pm$3.8 & 86.5$\pm$3.7 & 31.5$\pm$3.2 & 70.0$\pm$0.8  \\
\hline          
\multirow{10}{0.5cm}[0.23cm]{\rotatebox{90}{\parbox{1.9cm}{Semi-supervised \\DA methods}}}
& JointDomain & 26.2$\pm$2.7 & 98.6$\pm$0.4 & 35.3$\pm$2.0 & 50.8$\pm$2.5  \\
& DACoM \cite{li2018semi} & 59.5$\pm$4.7 & 82.8$\pm$2.7 & 28.4$\pm$2.6 & 69.2$\pm$2.2  \\
& CDFSB \cite{chen2019a} & 33.2$\pm$1.9 & 94.3$\pm$0.5 & 29.4$\pm$1.7 & 55.8$\pm$1.6  \\
& CDFSB++ \cite{chen2019a} & 47.1$\pm$8.8 & 93.7$\pm$1.9 & 37.3$\pm$1.1 & 66.1$\pm$5.6  \\
& MME \cite{saito2019semi} & 56.5$\pm$3.4 & 80.8$\pm$4.4 & 27.1$\pm$2.6 & 66.7$\pm$2.0  \\
& LFT \cite{Tseng2020Cross-Domain} & 64.2$\pm$4.2 & 84.6$\pm$6.7 & 32.2$\pm$4.5 & 73.5$\pm$0.5  \\
& CCSL \cite{wang2020contrastive} & 54.6$\pm$4.6 & 85.7$\pm$2.9 & 29.7$\pm$2.0 & 67.4$\pm$2.1  \\
& \cellcolor{gray!20}\textbf{FLARE} & \cellcolor{gray!20}\textbf{66.1$\pm$2.3} & \cellcolor{gray!20}87.4$\pm$2.3 & \cellcolor{gray!20}\textbf{37.4$\pm$2.5} & \cellcolor{gray!20}\textbf{75.7$\pm$0.8}  \\
\hline
\end{tabular}
\end{center}
\end{table}

\subsubsection{Transfer task from private dataset to public dataset}
Table \ref{tab:Comparisons_SOTA_from_private_to_public} reports the results on transfer task (S4,S1)$\rightarrow$S4. Again, to demonstrate the effect of knowledge transfer, we compare DA methods with target-only learning methods. It can be observed that most target-only learning methods exhibit better performance by the specificity measure, but poorer by the sensitivity measure compared with DA methods. This can be attributed to the lack of adequate positive training samples in the target domain. By transferring discriminant knowledge from Site-1, FLARE obtains the best results for sensitivity, ${\rm F1}$, and ${\rm G}{\text -}{\rm mean}$ measures under both balanced and imbalanced settings. Moreover, compared with the results under balanced setting, all DA methods achieve poorer performance in terms of ${\rm F1}$ measure under imbalanced setting, which is mainly due to the difference in dataset partition. Specifically, the two classes (mild versus severe) differ significantly in sample sizes (566 versus 32) under imbalanced setting. This leads to a large number of false positives when predicting for a difficult classification task. Hence, the models obtain a poor precision and a consequent poor ${\rm F1}$ measure.

\subsection{Ablation Study}
\label{subsec:Ablation Study}

In this section, we evaluate the effectiveness of each component in the FLARE method. Specifically, we implement our FLARE method on task (S1,S2)$\rightarrow$S1 under balanced setting by removing each proposed piece and replacing it with related components. In addition, we show the impact of varying each hyper-parameter.

\subsubsection{Effectiveness of the SCBS strategy}
To validate the effectiveness of the SCBS strategy, we remove it from the training regime, and optimize the model based on the raw class-imbalanced data or other imbalanced learning strategies, e.g., random oversampling, synthetic minority oversampling technique (SMOTE) \cite{chawla2002smote}, adaptive synthetic sampling approach (ADASYN) \cite{he2008adasyn}, integrations of SMOTE with edited nearest neighbor (SMOTE+ENN) \cite{batista2004study}, SMOTE with Tomek links (SMOTE+Tomek) \cite{batista2004study}, cRT \cite{Kang2020Decoupling}, focal loss \cite{lin2017focal}, and CR\cite{yang2019self}. The classification results in Table \ref{tab:Effectiveness_of_SCBS} show that all strategies boost the performance of the baseline. It can be observed that the simple random oversampling strategy improves the FLARE method by a large margin in ${\rm F1}$ and ${\rm G}{\text -}{\rm mean}$ (compared with baseline strategy). However, the improvement by the recently proposed focal loss \cite{lin2017focal} is smaller than the sampling strategies. We think the reason is that there is large intra-class heterogeneity among the CT images. Thus, it is difficult for an imbalanced learning strategy which only focuses on poorly-predicted class mining to achieve comparable performance. In the SCBS strategy, we first construct a class-balanced dataset and learn discriminant knowledge progressively from easy to hard classes. Hence the SCBS obtains an overall improvement.

\begin{table}
\caption{Performance of FLARE using various imbalanced learning strategies. $*$ means a statistically significant difference between the performances of the SCBS and those of the referred strategies.}
\label{tab:Effectiveness_of_SCBS}
\renewcommand\tabcolsep{2.5pt}
\begin{center}
\begin{tabular}{l|| cccc}
\hline
Strategies & \multicolumn{1}{c}{${\rm SEN}$} & \multicolumn{1}{c}{${\rm SPE}$} & \multicolumn{1}{c}{${\rm F1}$} & \multicolumn{1}{c}{${\rm G}{\text -}{\rm mean}$} \\
\hline
None (baseline) & 54.5$\pm$5.4$^*$ & 83.4$\pm$3.5$^*$ & 62.3$\pm$3.9$^*$ & 65.9$\pm$2.9$^*$  \\
Random oversampling & 79.1$\pm$1.0{\tiny~~} & 92.8$\pm$0.9{\tiny~~} & 84.9$\pm$0.5$^*$ & 85.6$\pm$0.5$^*$  \\
SMOTE \cite{chawla2002smote} & 74.9$\pm$1.2$^*$ & 91.5$\pm$1.4{\tiny~~} & 81.7$\pm$0.4$^*$ & 82.7$\pm$0.4$^*$  \\
Borderline1-SMOTE \cite{han2005borderline} & 73.9$\pm$1.2$^*$ & 91.0$\pm$1.3{\tiny~~} & 80.8$\pm$0.7$^*$ & 81.9$\pm$0.7$^*$  \\
Borderline2-SMOTE \cite{han2005borderline} & 71.8$\pm$1.5$^*$ & 90.5$\pm$1.8{\tiny~~} & 79.2$\pm$0.7$^*$ & 80.5$\pm$0.7$^*$  \\
ADASYN \cite{he2008adasyn} & 75.6$\pm$1.7{\tiny~~} & 91.3$\pm$1.2{\tiny~~} & 82.0$\pm$0.8$^*$ & 83.0$\pm$0.7$^*$  \\
SMOTE+ENN \cite{batista2004study} & 75.6$\pm$1.2$^*$ & 92.9$\pm$1.1{\tiny~~} & 82.7$\pm$0.5$^*$ & 83.7$\pm$0.4$^*$  \\
SMOTE+Tomek \cite{batista2004study} & 77.7$\pm$1.4{\tiny~~} & 89.2$\pm$1.4$^*$ & 82.4$\pm$0.6$^*$ & 83.1$\pm$0.5$^*$  \\
cRT \cite{Kang2020Decoupling} & 76.1$\pm$1.1$^*$ & 86.9$\pm$0.7$^*$ & 80.4$\pm$0.6$^*$ & 81.3$\pm$0.5$^*$  \\
Focal loss \cite{lin2017focal} & 50.3$\pm$3.1$^*$ & 88.5$\pm$1.9$^*$ & 61.7$\pm$2.4$^*$ & 66.2$\pm$1.8$^*$  \\
CR\cite{yang2019self} & 76.4$\pm$1.3$^*$ & \textbf{96.2$\pm$0.9}{\tiny~~} & 84.7$\pm$0.6{\tiny~~} & 85.7$\pm$0.5{\tiny~~}  \\
\rowcolor{gray!20}SCBS & \textbf{79.7$\pm$0.8}{\tiny~~} & 94.1$\pm$0.8{\tiny~~} & \textbf{85.8$\pm$0.4}{\tiny~~} & \textbf{86.5$\pm$0.4}{\tiny~~}  \\
\hline
\end{tabular}
\end{center}
\end{table}

\begin{figure}
\begin{center}
\includegraphics[width=1.00\linewidth]{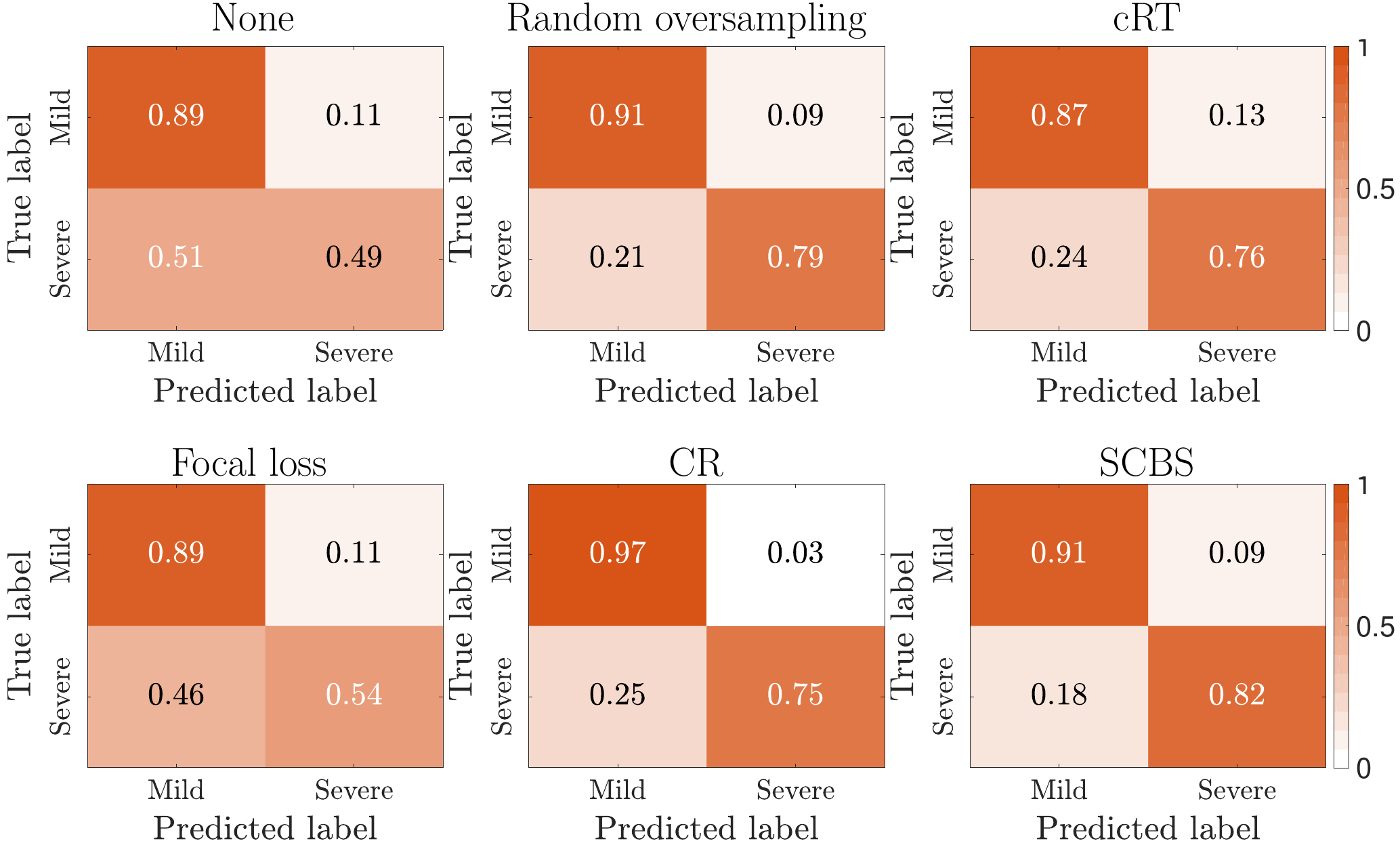} 
\end{center}
    \caption{{Confusion matrices of using various imbalanced learning strategies (None, random oversampling, cRT \cite{Kang2020Decoupling}, Focal loss \cite{lin2017focal}, CR \cite{yang2019self}, and SCBS) for the FLARE method.}}
    \label{fig:confu_Resize.}
\end{figure}

To check whether the SCBS strategy has statistically significant performance improvement compared with the referred strategies, we conduct paired Student’s t-test with a 0.05 significance level between these results. In each pair, we adopt the same dataset partition setting for two strategies to eliminate the possible impacts. According to Table \ref{tab:Effectiveness_of_SCBS}, most p-values are smaller than 0.05, demonstrating that SCBS has statistically significant performance improvement compared with most referred strategies. In addition, we report the confusion matrices for some strategies in Fig. \ref{fig:confu_Resize.}. We observe that the classification accuracy of the severe infection case improves by a large margin when using our SCBS strategy.

\subsubsection{Effectiveness of the domain translator and the prototype triplet loss}
\label{subsubsec:Effectiveness of domain translator and the prototype triplet loss}
We abbreviate the model without the domain translator, the prototype triplet loss, and their combination to FLARE (noD), FLARE (noP), and FLARE (noDP), respectively. We also replace the prototype triplet loss with the contrastive loss. The corresponding results are shown in Table \ref{tab:Effectiveness_of_component_DA_via_prototypeTriplet}. Comparing with these special cases, the improvement of FLARE (with selected hyper-parameter combination $\lambda_2 = 100$, $\alpha=0$) verifies the contributions of the domain translator and the prototype triplet loss in our method. In addition, we vary the value of the weight $\lambda_2$ and the margin $\alpha$. It can be observed that the performance of FLARE slightly fluctuates around the selected hyper-parameter combination.

\begin{table}[t]
\caption{Ablation results of FLARE method in terms of adaptive representation learning.}
\label{tab:Effectiveness_of_component_DA_via_prototypeTriplet}
\renewcommand\tabcolsep{1.7pt}
\begin{center}
\begin{tabular}{l|| cccc}
\hline
 & \multicolumn{1}{c}{${\rm SEN}$} & \multicolumn{1}{c}{${\rm SPE}$} & \multicolumn{1}{c}{${\rm F1}$} & \multicolumn{1}{c}{${\rm G}{\text -}{\rm mean}$} \\
\hline
FLARE (noDP) & 67.5$\pm$1.9 & 94.9$\pm$0.7 & 78.1$\pm$1.3 & 80.0$\pm$1.1 \\
FLARE (noDP)$+$contrastive loss & 69.5$\pm$1.0 & 93.2$\pm$1.1 & 78.8$\pm$0.5 & 80.4$\pm$0.4 \\
FLARE (noD) & 69.8$\pm$1.5 & 93.9$\pm$0.7 & 79.3$\pm$1.0 & 80.9$\pm$0.8 \\
FLARE (noP) & 77.5$\pm$1.4 & 92.7$\pm$1.6 & 83.9$\pm$0.4 & 84.7$\pm$0.4 \\
FLARE (noP)$+$contrastive loss & 79.0$\pm$1.1 & 91.3$\pm$1.1 & 84.1$\pm$0.5 & 84.9$\pm$0.4 \\
FLARE ($\lambda_2=10$, $\alpha=0$) & 78.9$\pm$1.3 & 91.7$\pm$1.4 & 84.3$\pm$0.5 & 85.0$\pm$0.4 \\
FLARE ($\lambda_2=50$, $\alpha=0$) & 76.6$\pm$1.3 & \textbf{95.1$\pm$1.0} & 84.3$\pm$0.5 & 85.2$\pm$0.4 \\
\rowcolor{gray!20}FLARE ($\lambda_2=100$, $\alpha=0$) & \textbf{79.7$\pm$0.8} & 94.1$\pm$0.8 & \textbf{85.8$\pm$0.4} & \textbf{86.5$\pm$0.4} \\
FLARE ($\lambda_2=500$, $\alpha=0$) & 79.0$\pm$1.2 & 92.3$\pm$0.8 & 84.6$\pm$0.5 & 85.3$\pm$0.4 \\
FLARE ($\lambda_2=1000$, $\alpha=0$) & 78.0$\pm$1.8 & 91.1$\pm$1.3 & 83.4$\pm$0.8 & 84.2$\pm$0.6 \\
FLARE ($\lambda_2=100$, $\alpha=0.4$) & 78.6$\pm$1.0 & 93.6$\pm$0.9 & 85.0$\pm$0.5 & 85.8$\pm$0.4 \\
FLARE ($\lambda_2=100$, $\alpha=0.8$) & 79.6$\pm$1.0 & 92.2$\pm$0.8 & 84.9$\pm$0.4 & 85.6$\pm$0.3 \\
FLARE ($\lambda_2=100$, $\alpha=1.2$) & 79.1$\pm$0.9 & 92.1$\pm$0.9 & 84.6$\pm$0.5 & 85.3$\pm$0.5 \\
FLARE ($\lambda_2=100$, $\alpha=1.6$) & 79.4$\pm$1.6 & 91.8$\pm$1.7 & 84.6$\pm$0.7 & 85.2$\pm$0.6 \\
FLARE ($\lambda_2=100$, $\alpha=2.0$) & 78.9$\pm$0.8 & 92.0$\pm$0.7 & 84.4$\pm$0.4 & 85.2$\pm$0.3 \\
\hline
\end{tabular}
\end{center}
\end{table}

\begin{table}[t]
\caption{{Ablation results of FLARE method in terms of discriminant learning.}}
\label{tab:Effectiveness_of_component_CMMD}
\renewcommand\tabcolsep{4pt}
\begin{center}
\begin{tabular}{l|| cccc}
\hline
 & \multicolumn{1}{c}{${\rm SEN}$} & \multicolumn{1}{c}{${\rm SPE}$} & \multicolumn{1}{c}{${\rm F1}$} & \multicolumn{1}{c}{${\rm G}{\text -}{\rm mean}$} \\
\hline
FLARE (noC)$+$CE loss & 79.5$\pm$1.1 & 82.4$\pm$1.1 & 80.7$\pm$0.5 & 80.9$\pm$0.5 \\
FLARE ($\lambda_1=0.1$) & 79.2$\pm$1.2 & 91.1$\pm$1.0 & 84.2$\pm$0.6 & 84.9$\pm$0.6 \\
FLARE ($\lambda_1=0.5$) & 79.2$\pm$1.1 & 92.1$\pm$0.6 & 84.7$\pm$0.7 & 85.4$\pm$0.6 \\
\rowcolor{gray!20}FLARE ($\lambda_1=1$) & 79.7$\pm$0.8 & \textbf{94.1$\pm$0.8} & \textbf{85.8$\pm$0.4} & \textbf{86.5$\pm$0.4} \\
FLARE ($\lambda_1=10$) & 78.7$\pm$1.3 & 91.6$\pm$0.9 & 84.1$\pm$0.6 & 84.8$\pm$0.5 \\
FLARE ($\lambda_1=100$) & \textbf{80.0$\pm$0.9} & 90.0$\pm$0.9 & 84.2$\pm$0.6 & 84.9$\pm$0.5 \\
\hline
\end{tabular}
\end{center}
\end{table}

\subsubsection{Effectiveness of the CMMD loss}
To explore the contribution of the CMMD loss in discriminant learning, we replace it with a standard cross-entropy loss. Comparing the results of FLARE (noC)$+$CE loss and FLARE (with default hyper-parameter $\lambda_1=1$) in Table \ref{tab:Effectiveness_of_component_CMMD}, we observe that equipping the CMMD loss increases the overall performance by a large margin. This suggests that the CMMD loss successfully improves the discrimination power of the latent space. In addition, the experimental results of varying the weight of the CMMD loss $\lambda_1$ show that FLARE obtains slightly poorer performance around the selected hyper-parameter.

\subsubsection{Effectiveness of the multi-view reconstruction loss}
We investigate the importance of the multi-view reconstruction loss. The model without the reconstruction loss is abbreviated as FLARE (noR). We also replace the multi-view decoder with a naive decoder (i.e., a fully connected layer from the latent feature to the input layer). As can be observed from Table \ref{tab:Effectiveness_of_component_reconstruction}, FLARE (with selected hyper-parameter $\lambda_3=0.002$) consistently outperforms FLARE (noR) and its variant with a naive decoder in terms of all metrics. This demonstrates that the completeness enables the latent space to make full use of multi-view features, and thus improves the classification effect. Similarly, we vary the value of the weight $\lambda_3$ and observe that the performance of FLARE fluctuates around the selected hyper-parameter.

\begin{table}
\caption{{Ablation results of FLARE method in terms of multi-view representation learning.}}
\label{tab:Effectiveness_of_component_reconstruction}
\renewcommand\tabcolsep{2.5pt}
\begin{center}
\begin{tabular}{l|| cccc}
\hline
 & \multicolumn{1}{c}{${\rm SEN}$} & \multicolumn{1}{c}{${\rm SPE}$} & \multicolumn{1}{c}{${\rm F1}$} & \multicolumn{1}{c}{${\rm G}{\text -}{\rm mean}$} \\
\hline
FLARE (noR) & 78.6$\pm$1.3 & 92.8$\pm$1.4 & 84.6$\pm$0.3 & 85.3$\pm$0.3 \\
FLARE (noR)$+$naive decoder & 79.6$\pm$1.3 & 91.9$\pm$1.1 & 84.7$\pm$0.5 & 85.4$\pm$0.4 \\
FLARE ($\lambda_3=0.0002$) & 80.3$\pm$1.3 & 90.4$\pm$1.2 & 84.5$\pm$0.7 & 85.1$\pm$0.6 \\
FLARE ($\lambda_3=0.001$) & \textbf{80.5$\pm$1.1} & 91.2$\pm$1.0 & 85.0$\pm$0.4 & 85.7$\pm$0.4 \\
\rowcolor{gray!20}FLARE ($\lambda_3=0.002$) & 79.7$\pm$0.8 & \textbf{94.1$\pm$0.8} & \textbf{85.8$\pm$0.4} & \textbf{86.5$\pm$0.4} \\
FLARE ($\lambda_3=0.005$) & 78.9$\pm$0.9 & 92.8$\pm$0.9 & 84.8$\pm$0.6 & 85.5$\pm$0.5 \\
FLARE ($\lambda_3=0.02$) & 78.3$\pm$1.5 & 92.2$\pm$1.2 & 84.1$\pm$0.7 & 84.9$\pm$0.6 \\
\hline
\end{tabular}
\end{center}
\end{table}

\subsection{Performance Under Various Training Regimes}
\label{subsec:Performance Under Various Training Regimes}

\begin{figure}[t]
    \centering
    \subfigure[]{
        \label{fig:Per_tl}
        \includegraphics[width=0.15\textwidth]{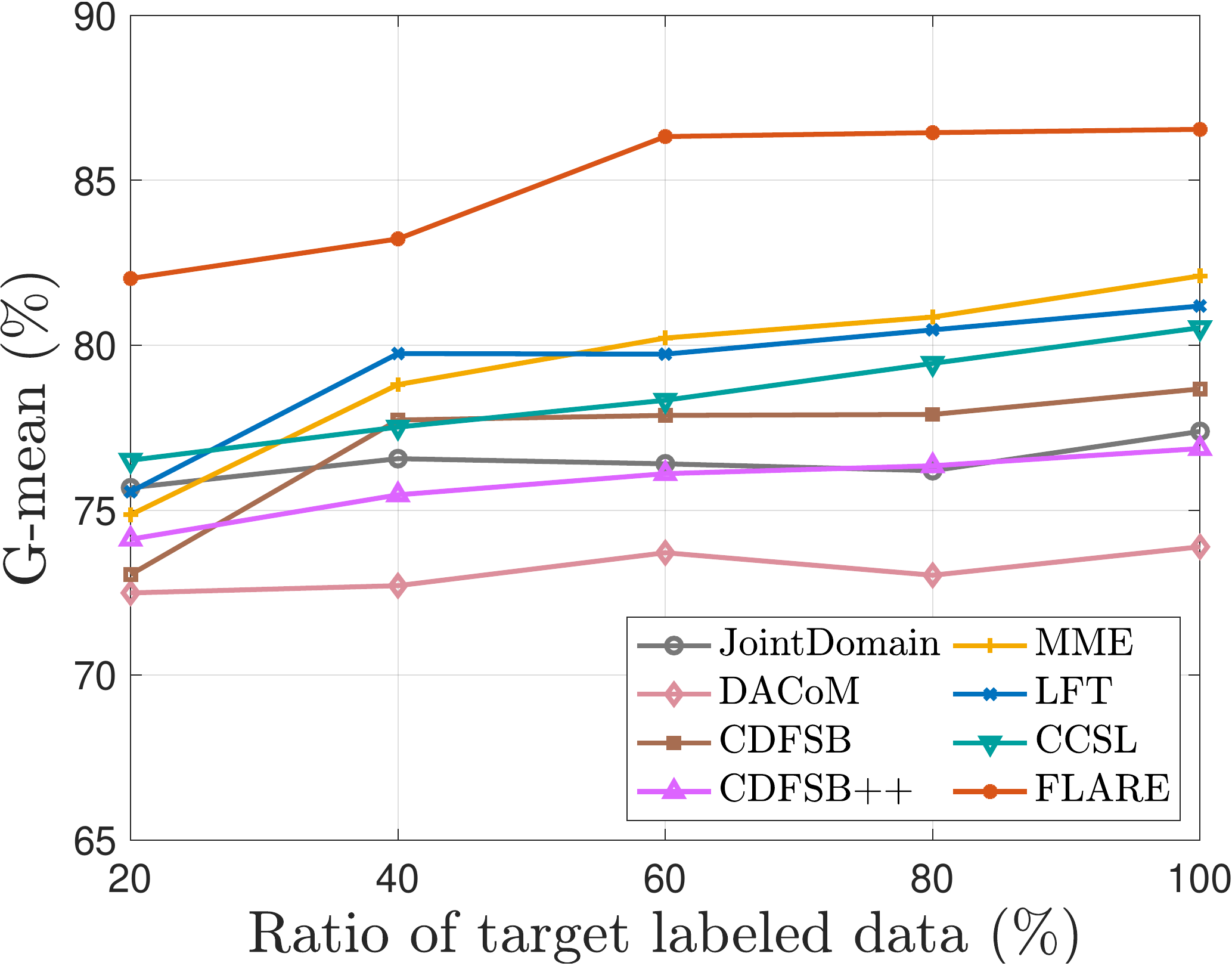}}
    \subfigure[]{
        \label{fig:Per_tu}
        \includegraphics[width=0.15\textwidth]{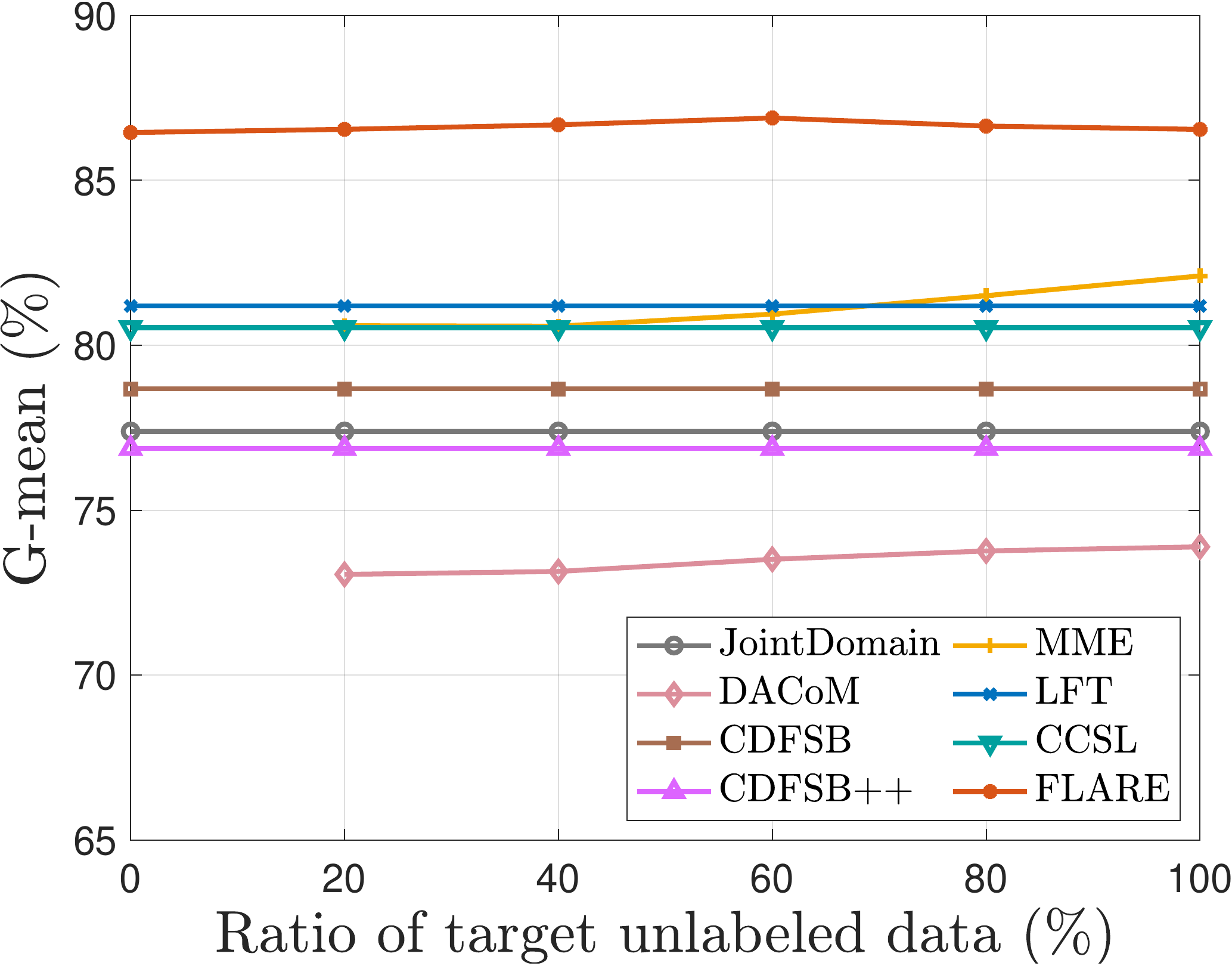}}
    \subfigure[]{
        \label{fig:Per_both}
        \includegraphics[width=0.15\textwidth]{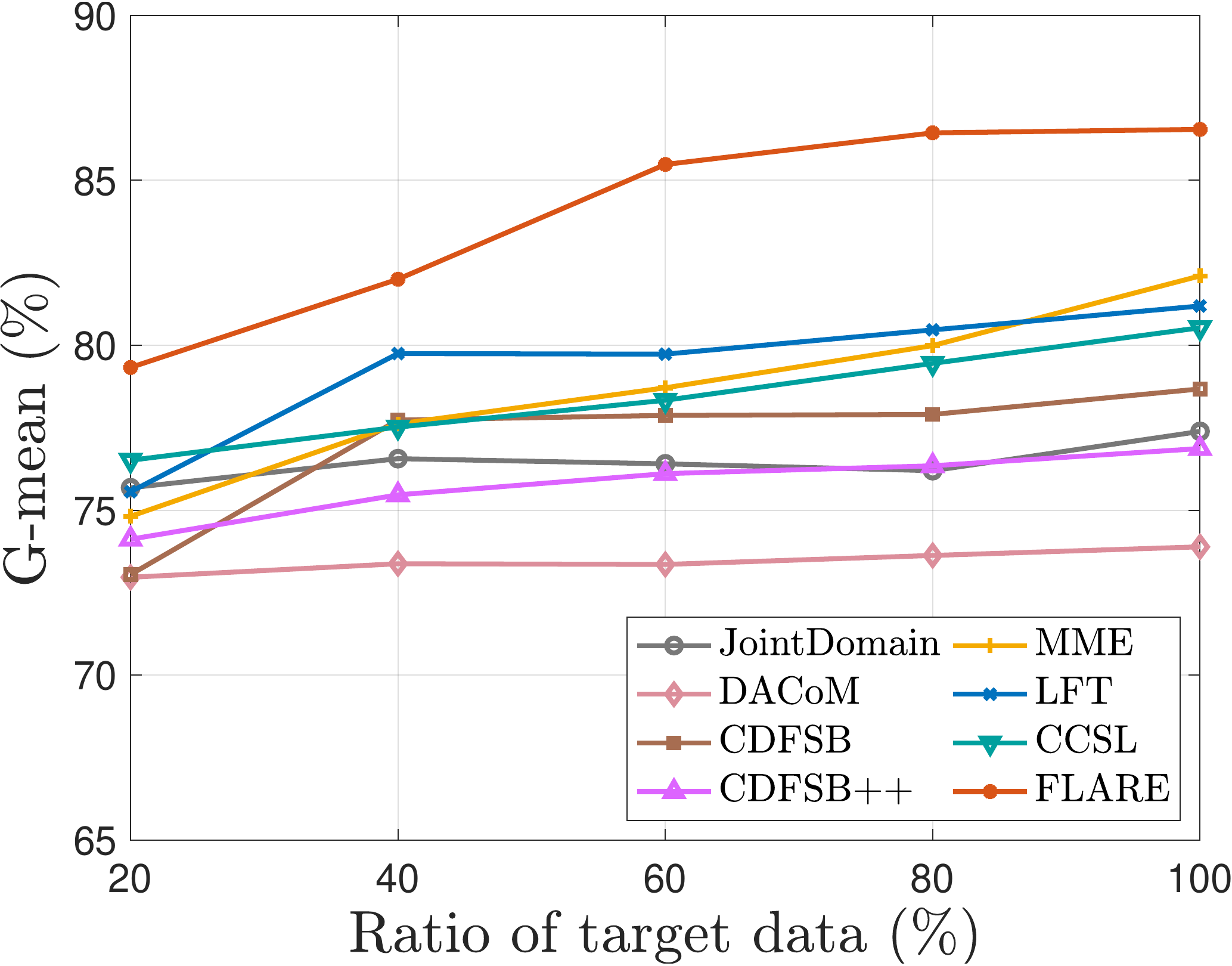}}
    \caption{Stability of FLARE versus other methods on the task (S1,S2)$\rightarrow$S1 with different ratios of (a) target labeled data, {(b) target unlabeled data, and (c) target data in the training phase.}}
    \label{fig:plot_performance_under_various_training_regimes.}
\end{figure}

To verify the robustness of our FLARE to reduced training data in the target domain, we train all the semi-supervised DA methods with three scenarios: 1) partial target labeled data, 2) partial target unlabeled data, and 3) partial target data. The plots of ${\rm G}{\text -}{\rm mean}$ scores on task (S1,S2)$\rightarrow$S1 under balanced setting are shown in Fig. \ref{fig:plot_performance_under_various_training_regimes.}. As illustrated in Fig. \ref{fig:Per_tl} and Fig. \ref{fig:Per_both}, along with a decrease in the ratio of the target (labeled) data, FLARE exhibits a graceful degradation and consistently outperforms other methods by a significant margin. Furthermore, when 60\% of the target (labeled) data is used, FLARE achieves comparable performance, which is even better than other methods with the entire target (labeled) dataset. We attribute this robustness to the prototype triplet loss specially designed for domain alignment. Specifically, when supervision information in the target domain is reduced, FLARE can still estimate the prototypes for all classes and project the data from the source domain around them for transferring discrimination knowledge. In addition, as shown in Fig. \ref{fig:Per_tu}, our FLARE method is insensitive to the ratio of target unlabeled data. However, this does not mean that we can ignore the contribution of the unlabeled data. In particular, when 20\% of target unlabeled data is used, FLARE obtains a ${\rm G}{\text -}{\rm mean}$ below 80\% (the far left of Fig. \ref{fig:Per_both}); after introducing more target unlabeled data (i.e., all target unlabeled data is used), FLARE boosts the ${\rm G}{\text -}{\rm mean}$ to 82\% (the far left of Fig. \ref{fig:Per_tl}).

\subsection{Feature Visualization}
\label{subsec:Visualization}

\begin{figure}[t]
\begin{center}
\includegraphics[width=1.00\linewidth]{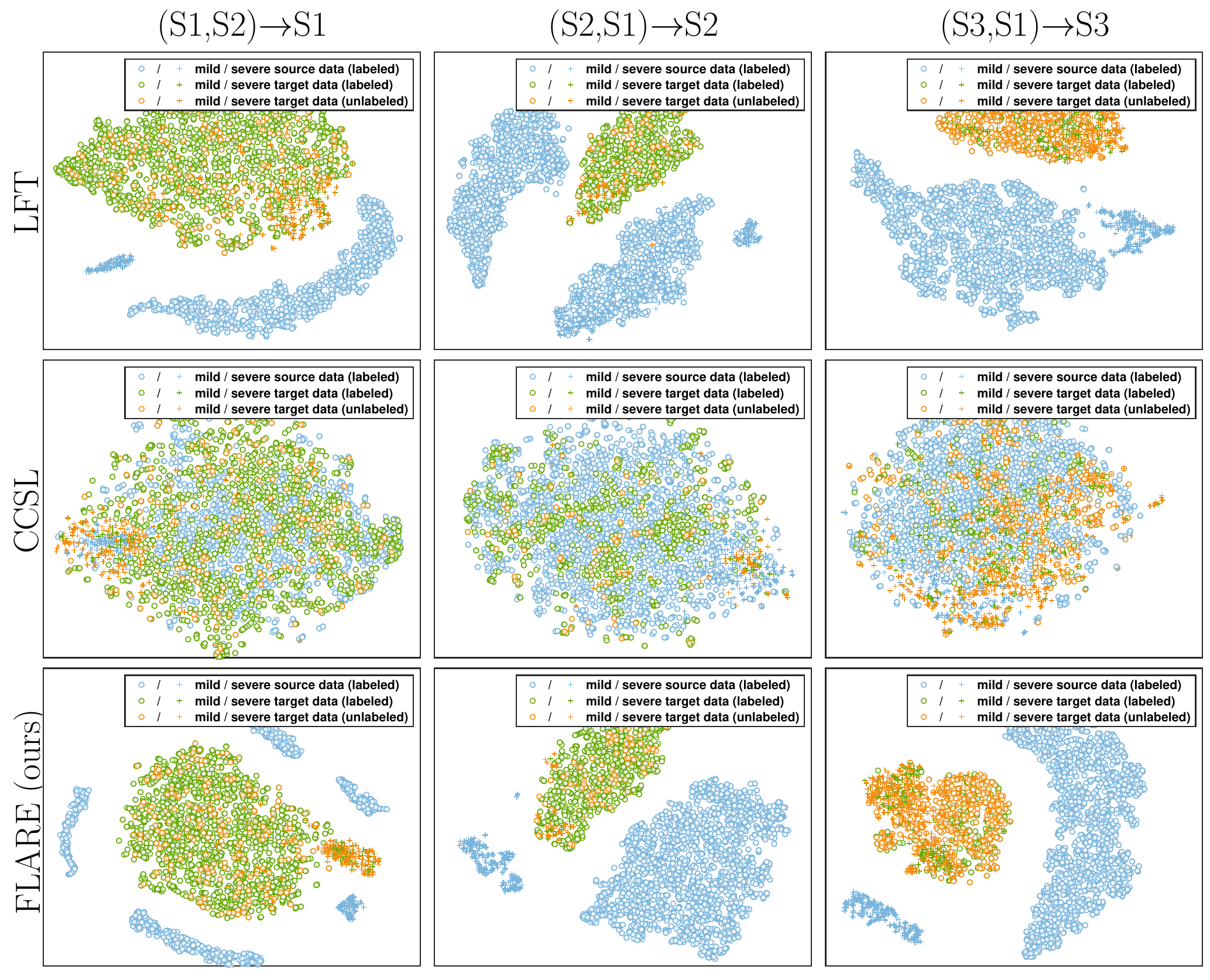}
\end{center}           
    \caption{Visualization of extracted features using t-SNE \cite{maaten2008visualizing}. Left to right: three DA tasks for classifying COVID-19 disease severity. Top to bottom: LFT \cite{Tseng2020Cross-Domain}, CCSL \cite{wang2020contrastive}, and our FLARE. FLARE is successful in projecting the source data around the target data and learning discriminant features. (Best viewed in color.)}
    \label{fig:tSNE.}
\end{figure}

We visualize the feature representations by t-SNE \cite{maaten2008visualizing} in Fig. \ref{fig:tSNE.} for tasks (S1,S2)$\rightarrow$S1, (S2,S1)$\rightarrow$S2, and (S3,S1)$\rightarrow$S3 under balanced setting. Two domains are aligned with CCSL \cite{wang2020contrastive}, but the classes are not well-discriminated. As for LFT \cite{Tseng2020Cross-Domain} and FLARE, they both adopt a domain transformation. LFT \cite{Tseng2020Cross-Domain} inserts the feature-wise transformation layers into the feature extractor to obtain various feature distributions. This operation improves the prediction ability of the model on target unlabeled samples to a certain extent, but lacks consideration of category information, as is shown in Fig. \ref{fig:tSNE.}. Different from LFT \cite{Tseng2020Cross-Domain}, FLARE uses a domain translator before the shared feature extractor, and then aligns two domains on the latent space via a prototype triplet loss. Fig. \ref{fig:tSNE.} illustrates that FLARE successfully projects the source data around the target data with the same category, and transfers the discrimination information simultaneously. Therefore, FLARE obtains larger class separability than LFT.

\begin{figure}[t]
\begin{center}
\includegraphics[width=1.00\linewidth]{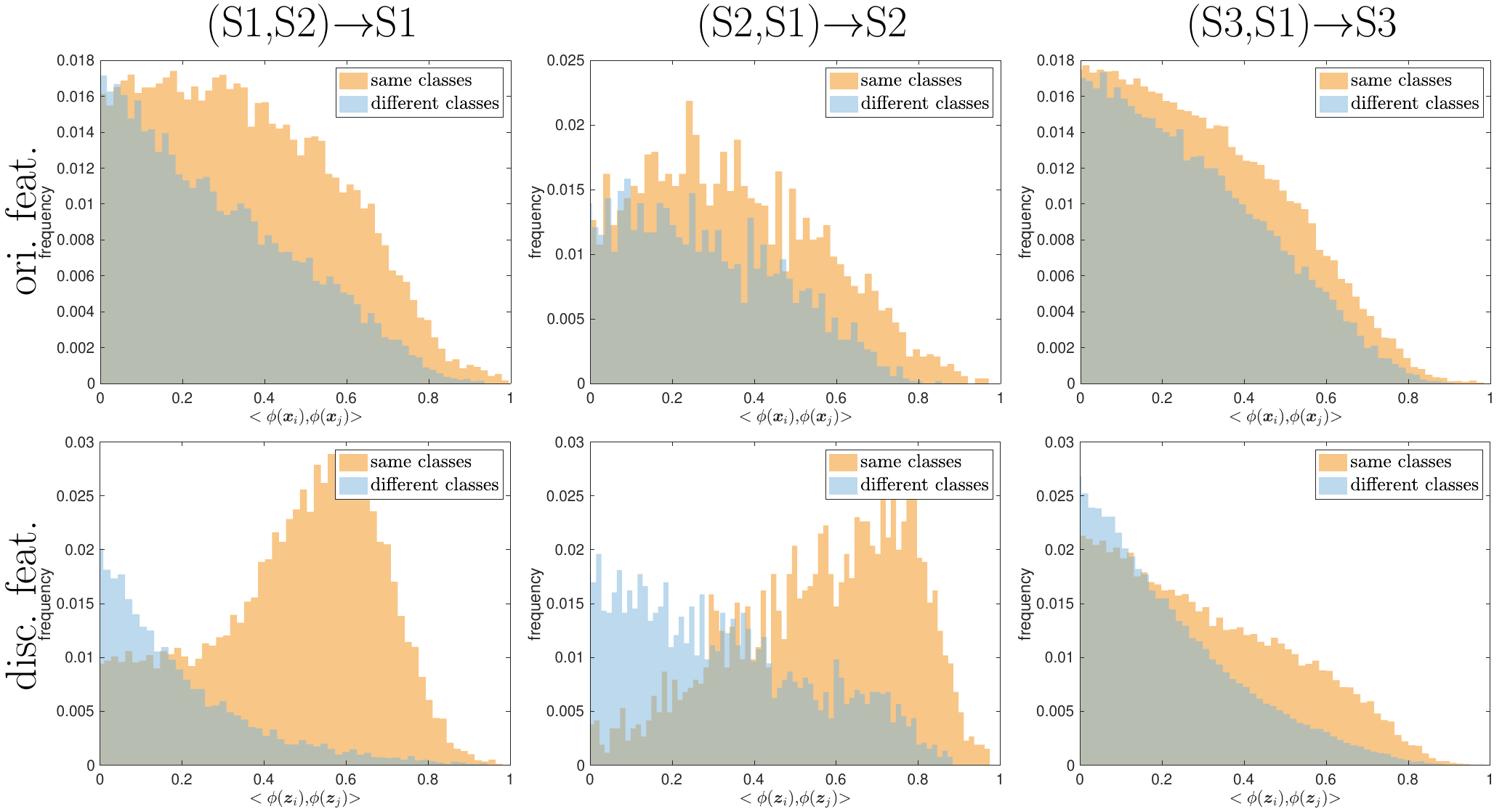}
\end{center}
    \caption{Similarity distribution of features. Left to right: three DA tasks for COVID-19 severity assessment. Top: original features of the target domain. Bottom: discriminative features extracted by our FLARE for the target domain. In each figure, we draw two batches of data with the same and different classes. Compared with the original features, the intra-class similarity of the features learned by FLARE becomes larger and the inter-class one becomes smaller. (Best viewed in color.)}
    \label{fig:hist of kernel value using our method and original data.}
\end{figure}

To further investigate the discriminant of features, we compare the similarities of features with the same class and different classes in Fig. \ref{fig:hist of kernel value using our method and original data.}. In the top line of Fig. \ref{fig:hist of kernel value using our method and original data.}, we display the histograms of kernel values for the raw input features. Results on three DA tasks show that the similarities between all samples are relatively low, and two distributions (orange and blue) have no significant difference. Thus, a general kernel function on the raw samples cannot effectively reflect similarity.

\begin{table}
\caption{Comparisons against competing methods on multiple source domain adaptation task.}
\label{tab:multi-source using all labeled data}
\renewcommand\tabcolsep{4pt}
\begin{center}
\begin{tabular}{c| c|| cccc}
\hline
Standards & Methods & ${\rm SEN}$ & ${\rm SPE}$ & ${\rm F1}$ & ${\rm G}{\text -}{\rm mean}$ \\
\hline
\multirow{4}{1.1cm}{Single-\\best \\ domain \\ adaptation}
& MME \cite{saito2019semi} & 74.7$\pm$0.6 & 90.4$\pm$1.2 & 81.0$\pm$0.5 & 82.1$\pm$0.5 \\
& LFT \cite{Tseng2020Cross-Domain} & 73.2$\pm$1.8 & 90.5$\pm$2.2 & 80.1$\pm$0.8 & 81.2$\pm$0.7 \\
& CCSL \cite{wang2020contrastive} & 78.4$\pm$1.5 & 84.1$\pm$0.8 & 80.7$\pm$0.9 & 81.2$\pm$0.8 \\
& \textbf{FLARE} & 79.7$\pm$0.8 & \textbf{94.1$\pm$0.8} & 85.8$\pm$0.4 & 86.5$\pm$0.4 \\
\hline
\multirow{4}{1.1cm}{Source \\ combine \\ domain \\ adaptation}
& MME \cite{saito2019semi} & 76.2$\pm$1.1 & 84.1$\pm$0.6 & 79.3$\pm$0.7 & 80.0$\pm$0.6  \\
& LFT \cite{Tseng2020Cross-Domain} &  72.9$\pm$1.1 & 88.0$\pm$2.3 & 78.9$\pm$0.8 & 80.0$\pm$0.9 \\
& CCSL \cite{wang2020contrastive} & 77.8$\pm$1.3 & 86.3$\pm$0.9 & 81.2$\pm$0.7 & 81.9$\pm$0.7 \\
& \textbf{FLARE} & 80.3$\pm$1.7 & 87.6$\pm$1.9 & 83.3$\pm$0.7 & 83.7$\pm$0.7  \\
\hline
\multirow{4}{1.1cm}{Multi-\\source \\ domain \\ adaptation}
& DCTN \cite{xu2018deep} & 76.9$\pm$0.9 & 80.6$\pm$1.1 & 78.4$\pm$0.4 & 78.7$\pm$0.4  \\
& M$^3$SDA \cite{peng2019moment} & 74.5$\pm$1.4 & 86.6$\pm$2.2 & 79.3$\pm$0.6 & 80.2$\pm$0.6 \\
& MSCAN \cite{kang2020contrastive} & 79.2$\pm$1.0 & 88.1$\pm$1.1 & 82.9$\pm$0.5 & 83.5$\pm$0.5  \\
& \cellcolor{gray!20}\textbf{M-FLARE} & \cellcolor{gray!20}\textbf{83.3$\pm$0.5} & \cellcolor{gray!20}91.5$\pm$0.8 & \cellcolor{gray!20}\textbf{86.9$\pm$0.3} & \cellcolor{gray!20}\textbf{87.3$\pm$0.3}  \\
\hline
\end{tabular}
\end{center}
\end{table}

In the bottom line of Fig. \ref{fig:hist of kernel value using our method and original data.}, we plot the histograms of kernel values for the representations learned by FLARE. On the DA task (S1,S2)$\rightarrow$S1, the similarities of samples with different classes mainly concentrate below 0.4, whereas those with the same class concentrate between 0.4 and 0.8. In addition, on the task (S2,S1)$\rightarrow$S2, the kernel values of samples with different classes mainly concentrate below 0.6, whereas those with the same class are concentrated in the range of 0.5 to 0.9. On these two DA tasks, a clear distinction between two distributions means that the representations learned by FLARE have better discriminant. On the last DA task, (S3,S1)$\rightarrow$S3, the distinction between the two distributions is small. We infer the reason is that there is large within-class heterogeneity in the CT images. In particular, we find that many samples on Site-3 are critical severe or even dead cases. The presence of coexisting illnesses is common among these patients (also observed in \cite{guan2020clinical}), which results in small within-class similarities.

\subsection{{Evaluation on Multi-source Domain Adaptation}}
\label{subsec:Multiple Source Domains}

\begin{figure}
\begin{center}
\includegraphics[width=0.72\linewidth]{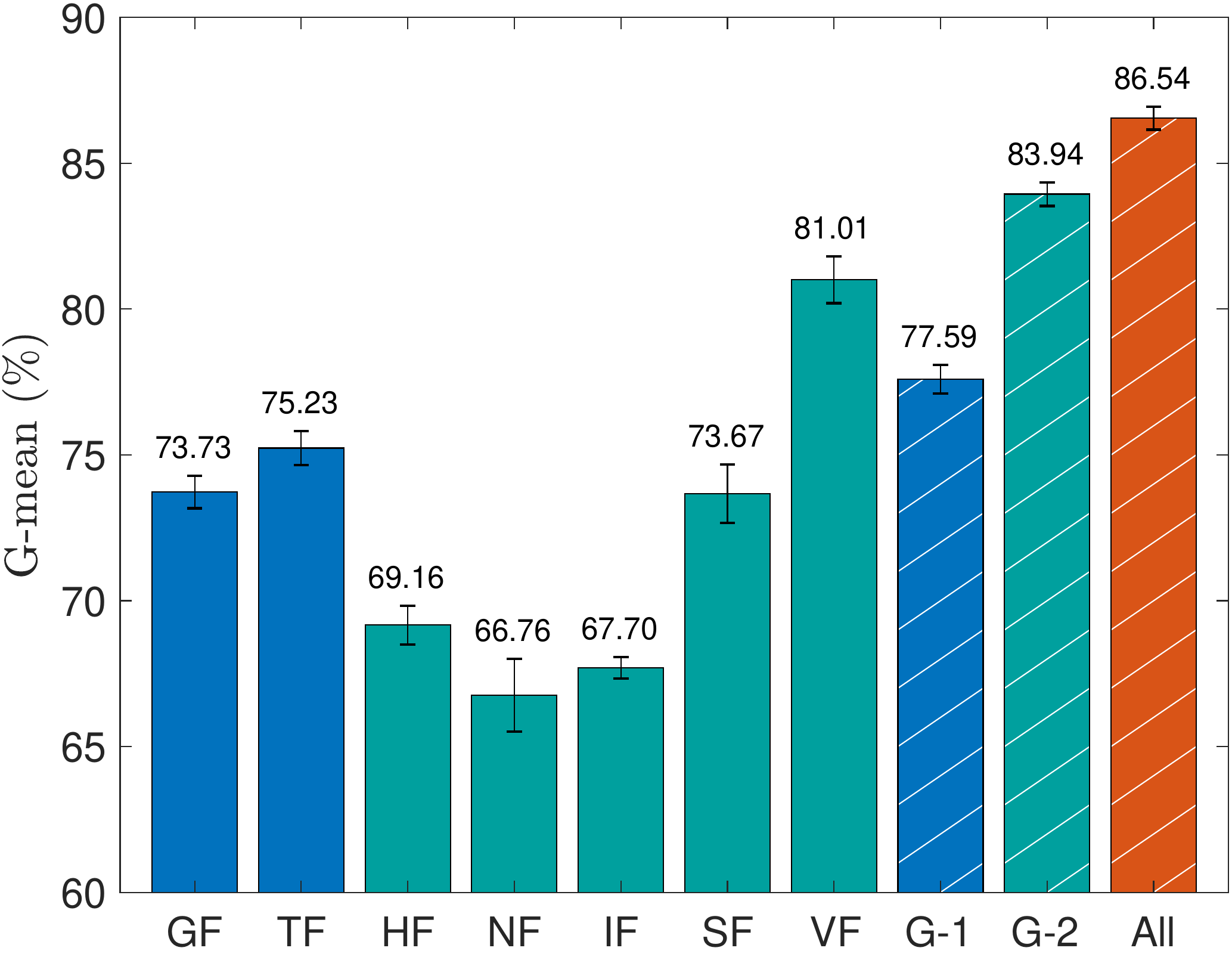}
\end{center}
   \caption{{Performance of FLARE method using different types of features. GF, TF, HF, NF, IF, SF, and VF denote gray features, texture features, histogram features, number features, intensity features, surface features, and volume features, respectively. These features are further divided into two groups: G-1 (including GF and TF) and G-2 (including the rest). ``All'' represents all types of features.}}
\label{fig:Gmean of different types of features}
\end{figure}

Table \ref{tab:multi-source using all labeled data} shows the comparisons of M-FLARE with some recent methods on the multi-source DA task (S1,S2,S3)$\rightarrow$S1. Following \cite{xu2018deep, peng2019moment, kang2020contrastive}, we introduce two standards: 1) single-best domain adaptation, which reports the single-source adaptation result best-performing in the test set, and 2) source combine domain adaptation, which reports the single-source adaptation result by transferring the combination of multiple source domains to a target domain. The first standard evaluates whether the single-source adaptation performance can be improved by introducing other source domains; the second testifies the need of exploiting a multi-source DA method. From the experimental results in Table \ref{tab:multi-source using all labeled data}, we have three observations. Firstly, the single-best FLARE method obtains much better classification performance than other single-source methods and most multi-source methods. This validates the effectiveness of FLARE. Secondly, compared with single-best DA methods, most source combine DA methods obtain poorer results, demonstrating that domain shifts between source domains should not be neglected. Thirdly, M-FLARE boosts the performance of single-best and source combine FLARE to 87.3\% in terms of ${\rm G}{\text -}{\rm mean}$, which validates that introducing additional source domains and adopting appropriate transfer method can improve the performance on the target domain.

\subsection{Discrimination Power and Complementarity of Different Types of Features}
\label{subsec:Discrimination power and complementarity of different types of features}
To quantitatively investigate the importance and the complementarity of different types of features for assessing COVID-19 severity, we conduct experiments on each type of feature and concatenated features with our FLARE method. Fig. \ref{fig:Gmean of different types of features} presents the corresponding results on task (S1,S2)$\rightarrow$S1 under balanced setting. We can observe that the severity assessment performance varies considerably between different types of features. For instance, number features (NF) and intensity features (IF) have poor discrimination ability. One reason may be that the number of lesions and the intensity in lung are not quite different between patients with various degrees of COVID-19 infections. In addition, among the seven types of features, volume features (VF) have significant discrimination ability. This conforms to the clinical diagnosis. For example, pulmonary parenchyma lesion volume in chest CT plays a key role in COVID-19 disease grading and decision of treatment planning~\cite{morozov2020mosmeddata}. Although different types of features have different discrimination power, there is complementarity among them. As shown in Fig. \ref{fig:Gmean of different types of features}, FLARE based on concatenated features (i.e., G-1, G-2, and All) obtains larger ${\rm G}{\text -}{\rm mean}$ than those on a single type of feature, which sufficiently supports the need for jointly using multiple types of features.

\section{Discussion}
\label{sec:Discussion}

In this section, we discuss about the robustness and the time complexity of our FALRE method, and some future works.

\subsection{Robustness}

\subsubsection{Robustness to the sample size}
\label{subsubsec:Discussion-Robustness to the amounts of labeled samples}
The proposed FLARE method is effective for transferring discriminant information from a clinical site (source domain) to another site (target domain). This is verified in extensive experiments on transfer tasks between different datasets. Let us take the DA task (S1,S2)$\rightarrow$S1 with imbalanced setting as an example. The source domain (Site-2) and the target domain (Site-1) contain respectively 1,245 and 818 samples. From the results in Table \ref{tab:Comparisons_SOTA_1}, we observe that the best target-only supervised learning method obtains 79.6\% in terms of ${\rm G}{\text -}{\rm mean}$, whereas the best semi-supervised DA method boosts the performance to 86.3\%. We contribute this to the transfer learning in the semi-supervised DA method. Note that different sites have different amounts of labeled samples, and thus the DA tasks between these sites (e.g., (S1,S2)$\rightarrow$S1 and (S2,S1)$\rightarrow$S2) have different ratios of source to target sample sizes. For full comparisons, we report the results on all DA tasks between private datasets in Table \ref{tab:Comparisons_SOTA_1} (Section \ref{subsec:Comparison with SOTA}). In addition, to further investigate the effects of sample sizes, we implement experiments on two additional scenarios: 1) multi-source domain adaptation where there are multiple source domains (Section \ref{subsec:Multiple Source Domains}), and 2) various training regimes where partial target labeled data is used (Section \ref{subsec:Performance Under Various Training Regimes}). The corresponding experimental results confirm that our FLARE is robust to the sample size. In addition, introducing more discriminant knowledge and adopting appropriate DA methods are beneficial for the improvement of classification performance on the target domain.

\subsubsection{Robustness to the image quality and resolution}
\label{subsubsec:Discussion-Robustness to the Image Quality and Resolution}
In this study, the CT images are collected from several CT scanners, including uCT 780 from UIH, Optima CT520, Discovery CT750, LightSpeed 16 from GE, Aquilion ONE from Toshiba, SOMATOMForce from Siemens, and SCENARIA from Hitachi. Hence the image qualities of these CT samples are varied. In clinical practice, due to the consideration of radiation doses and the difference in the capacity of CT scanners, CT slice thickness is various within and between hospitals. For instance, in the collected private datasets, reconstructed CT thickness ranges from 0.625 to 2 mm; in the public MosMedData, the thickness is 8 mm. For normalization processing, we resample all CT images into 1.5 mm isotropic resolution as \cite{shi2021large}. Although the image quality and CT thickness are varied, extensive experiments on different sites show that our FLARE method obtains compelling results compared with the competitors. This demonstrates the robustness of the proposed framework to the image quality and resolution.

\subsection{Time Complexity}
The FLARE method contains three major stages: 1) multi-view feature extraction; 2) parameter optimization of the FCN; 3) calculation of the objective function. In the first stage, we use a pre-trained VB-Net \cite{shan2021abnormal} to segment the infected lesions and lung fields, and then extract handcrafted features for all images. We conduct this feature extraction operation only once and fix the multi-view features during the algorithm training phase. In the second stage, the domain translator, the feature extractor, and the classifier form an FCN at a cost of $\mathcal{O}(\sum_{l=1}^{d_f} w_{l-1} w_{l})$ where $d_f$ is the number of layers, and $w_{l}$ is the dimension of the $l$-th layer. The third stage contains the calculations of the CMMD loss, the prototype triplet loss, and the reconstruction loss. These loss terms have time complexity $\mathcal{O}(n_b^3)$, $\mathcal{O}(n_b C)$, and $\mathcal{O}(\sum_{v=1}^{V} w_{z} w_{x}^{(v)})$, respectively, where $n_b$ is the batch size ($n_b = 100$ in the experiments when adopting the mini-batch training strategy), $w_{z}$ and $w_{x}^{(v)}$ are the dimensions of latent feature and the $v$-th view of input data. In our experiments, it takes an average of 40.0 seconds to segment the infections and lung fields and extract handcrafted features for a CT image on two Nvidia Quadro RTX 4000 graphics cards. In the algorithm training phase, it takes 1.6 seconds per iteration. During the testing phase, it only takes 0.6 milliseconds to assess the severity for a sample on an Nvidia GeForce GTX 1080 Ti graphics card.

With regard to probable improvement of efficiency, we can implement multi-threading and extract features for multiple CT images. When using a large batch size in the algorithm training phase, we can use the memory efficient kernel approximation~\cite{si2017memory}, which reduces the complexity of the CMMD loss from $\mathcal{O}(n_b^3)$ to $\mathcal{O}(n_b k^2 + k^3)$ with rank $k\ll n_b$.

\subsection{Future Works}
Recently, many deep learning methods based on original medical images have made great progress in the performance of medical image analysis. Nevertheless, we explore the COVID-19 severity assessment problem using multi-view features in this paper. One of the main reasons why we adopt handcrafted features is the particularity of the task. For instance, when detecting COVID-19 from non-COVID-19 pneumonia or non-pneumonia diseases, deep learning systems are proposed to detect some typical signs of infection (e.g., ground-glass opacity (GGO) and consolidation) in a qualitative evaluation way \cite{song2021deep, yazdani2020covid, ko2020covid}. However, when assessing COVID-19 severity, the quantitative indicators (e.g., the number and size of GGOs, the volume of infection) are very important \cite{huang2020clinical, morozov2020mosmeddata}. Hence, we extract multi-view features based on the lesion region and subsequently propose a DA method. However, it is undeniable that handcrafted features even with expert knowledge may miss some discriminative information. An interesting research direction is to take both original medical images and multi-view features into account. More rigorous analysis on this direction will be our future work.

Imbalanced learning is an important and challenging problem in many classification tasks, especially in the field of medicine. To learn a fair prediction for all classes, we propose an imbalanced learning strategy, named SCBS. In fact, generative model is also powerful and widely-explored to handle data imbalance issues. For example, in a pioneering work \cite{wang2021towards}, the feature space of minor class in the source domain is augmented by synthesizing fake data through a conditional generative adversarial network; experimental results show that this method achieves significant improvement in both minor and overall classification accuracy compared with some baselines. Inspired by this work, a promising direction is to augment the sample sizes of minor and hard classes by generating new data instead of oversampling data in the SCBS strategy. Meanwhile, the complexity of the synthetic sampling algorithm will be studied.

\section{Conclusion}
\label{sec:Conclusion}
In this paper, we propose a novel DA method, named FLARE, for the severity assessment of COVID-19. In particular, it tackles the class-imbalance problem and domain discrepancy among cross-site transfer learning. The key components in FLARE include an SCBS strategy and a desirable representation learning framework. The SCBS is specially designed for relieving the class-imbalance problem and improves the classification performance on poorly-predicted classes. In the representation learning, we pursue a latent space containing three properties, domain-transferability, discriminant, and completeness. For domain-transferability, we propose a domain translator and align the heterogeneous data via a prototype triplet loss in a hyper-sphere manifold. For discriminant and completeness, we adopt the CMMD loss and the multi-view reconstruction loss on the representations, respectively. Extensive experiments show that FLARE significantly outperforms some state-of-the-art methods.

The way to combine the dynamic changes of CT images during COVID-19 development to further improve the performance of the classification method will be our future work.

\bibliographystyle{IEEEtran}
\bibliography{FLARE_bib}

\begin{thebibliography}{10}
\providecommand{\url}[1]{#1}
\csname url@samestyle\endcsname
\providecommand{\newblock}{\relax}
\providecommand{\bibinfo}[2]{#2}
\providecommand{\BIBentrySTDinterwordspacing}{\spaceskip=0pt\relax}
\providecommand{\BIBentryALTinterwordstretchfactor}{4}
\providecommand{\BIBentryALTinterwordspacing}{\spaceskip=\fontdimen2\font plus
\BIBentryALTinterwordstretchfactor\fontdimen3\font minus
  \fontdimen4\font\relax}
\providecommand{\BIBforeignlanguage}[2]{{%
\expandafter\ifx\csname l@#1\endcsname\relax
\typeout{** WARNING: IEEEtran.bst: No hyphenation pattern has been}%
\typeout{** loaded for the language `#1'. Using the pattern for}%
\typeout{** the default language instead.}%
\else
\language=\csname l@#1\endcsname
\fi
#2}}
\providecommand{\BIBdecl}{\relax}
\BIBdecl

\bibitem{wang2020novel}
C.~Wang, P.~W. Horby, F.~G. Hayden, and G.~F. Gao, ``A novel coronavirus
  outbreak of global health concern,'' \emph{Lancet}, vol. 395, no. 10223, pp.
  470--473, 2020.

\bibitem{huang2020clinical}
C.~Huang \emph{et~al.}, ``Clinical features of patients infected with 2019
  novel coronavirus in wuhan, china,'' \emph{Lancet}, vol. 395, no. 10223, pp.
  497--506, 2020.

\bibitem{li2020early}
Q.~Li \emph{et~al.}, ``Early transmission dynamics in wuhan, china, of novel
  coronavirus--infected pneumonia,'' \emph{N. Engl. J. Med.}, vol. 382, no.~13,
  pp. 1199--1207, 2020.

\bibitem{chen2020epidemiological}
N.~Chen \emph{et~al.}, ``Epidemiological and clinical characteristics of 99
  cases of 2019 novel coronavirus pneumonia in wuhan, china: a descriptive
  study,'' \emph{Lancet}, vol. 395, no. 10223, pp. 507--513, 2020.

\bibitem{song2021deep}
Y.~Song \emph{et~al.}, ``Deep learning enables accurate diagnosis of novel
  coronavirus (covid-19) with ct images,'' \emph{IEEE-ACM Trans. Comput. Biol.
  Bioinform.}, 2021, DOI: 10.1109/TCBB.2021.3065361.

\bibitem{yazdani2020covid}
S.~Yazdani, S.~Minaee, R.~Kafieh, N.~Saeedizadeh, and M.~Sonka, ``Covid ct-net:
  Predicting covid-19 from chest ct images using attentional convolutional
  network,'' \emph{arXiv preprint arXiv:2009.05096}, 2020.

\bibitem{ko2020covid}
H.~Ko \emph{et~al.}, ``Covid-19 pneumonia diagnosis using a simple 2d deep
  learning framework with a single chest ct image: model development and
  validation,'' \emph{J. Med. Internet Res.}, vol.~22, no.~6, p. e19569, 2020.

\bibitem{saeedizadeh2021covid}
N.~Saeedizadeh, S.~Minaee, R.~Kafieh, S.~Yazdani, and M.~Sonka, ``Covid
  tv-unet: Segmenting covid-19 chest ct images using connectivity imposed
  u-net,'' \emph{Comput. Meth. Prog. Bio. Update}, vol.~1, p. 100007, 2021,
  DOI: 10.1016/j.cmpbup.2021.100007.

\bibitem{kang2020diagnosis}
H.~Kang \emph{et~al.}, ``Diagnosis of coronavirus disease 2019 (covid-19) with
  structured latent multi-view representation learning,'' \emph{IEEE Trans.
  Med. Imag.}, vol.~39, no.~8, pp. 2606--2614, 2020.

\bibitem{chassagnon2020ai}
G.~Chassagnon \emph{et~al.}, ``Ai-driven quantification, staging and outcome
  prediction of covid-19 pneumonia,'' \emph{Med. Image Anal.}, vol.~67, p.
  101860, 2021.

\bibitem{ouyang2020dual}
X.~Ouyang \emph{et~al.}, ``Dual-sampling attention network for diagnosis of
  covid-19 from community acquired pneumonia,'' \emph{IEEE Trans. Med. Imag.},
  vol.~39, no.~8, pp. 2595--2605, 2020.

\bibitem{minaee2020deep}
S.~Minaee, R.~Kafieh, M.~Sonka, S.~Yazdani, and G.~J. Soufi, ``Deep-covid:
  Predicting covid-19 from chest x-ray images using deep transfer learning,''
  \emph{Med. Image Anal.}, vol.~65, p. 101794, 2020.

\bibitem{chaganti2020automated}
S.~Chaganti \emph{et~al.}, ``Automated quantification of ct patterns associated
  with covid-19 from chest ct,'' \emph{Radiol. Artif. Intell.}, vol.~2, no.~4,
  p. e200048, 2020.

\bibitem{tang2021severity}
Z.~Tang \emph{et~al.}, ``Severity assessment of {COVID}-19 using {CT} image
  features and laboratory indices,'' \emph{Phys. Med. Biol.}, vol.~66, no.~3,
  p. 035015, 2021.

\bibitem{gibson2018inter}
E.~Gibson \emph{et~al.}, ``Inter-site variability in prostate segmentation
  accuracy using deep learning,'' in \emph{Proc. Med. Image Comput.
  Comput.-Assist. Intervent.}, 2018, pp. 506--514.

\bibitem{liu2020ms}
Q.~Liu, Q.~Dou, L.~Yu, and P.~A. Heng, ``Ms-net: Multi-site network for
  improving prostate segmentation with heterogeneous mri data,'' \emph{IEEE
  Trans. Med. Imag.}, vol.~39, no.~9, pp. 2713--2724, 2020.

\bibitem{zech2018variable}
J.~R. Zech, M.~A. Badgeley, M.~Liu, A.~B. Costa, J.~J. Titano, and E.~K.
  Oermann, ``Variable generalization performance of a deep learning model to
  detect pneumonia in chest radiographs: a cross-sectional study,'' \emph{PLoS
  Med.}, vol.~15, no.~11, pp. 1--17, 2018.

\bibitem{wang2020contrastive}
Z.~Wang, Q.~Liu, and Q.~Dou, ``Contrastive cross-site learning with redesigned
  net for covid-19 ct classification,'' \emph{IEEE J. Biomed. Health Inform.},
  vol.~24, no.~10, pp. 2806--2813, 2020.

\bibitem{verity2020estimates}
R.~Verity \emph{et~al.}, ``Estimates of the severity of coronavirus disease
  2019: a model-based analysis,'' \emph{Lancet Infect. Dis.}, vol.~20, no.~6,
  pp. 669--677, 2020.

\bibitem{li2020automated}
M.~D. Li \emph{et~al.}, ``Automated assessment of covid-19 pulmonary disease
  severity on chest radiographs using convolutional siamese neural networks,''
  \emph{Radiol. Artif. Intell.}, vol.~2, no.~4, p. e200079, 2020.

\bibitem{zhu2020joint}
X.~Zhu \emph{et~al.}, ``Joint prediction and time estimation of covid-19
  developing severe symptoms using chest ct scan,'' \emph{Med. Image Anal.},
  vol.~67, no.~8, p. 101824, 2020.

\bibitem{pan2010survey}
S.~J. Pan and Q.~Yang, ``A survey on transfer learning,'' \emph{IEEE Trans.
  Knowl. Data Eng.}, vol.~22, no.~10, pp. 1345--1359, 2010.

\bibitem{dong2020weakly}
J.~Dong, Y.~Cong, G.~Sun, Y.~Yang, X.~Xu, and Z.~Ding, ``Weakly-supervised
  cross-domain adaptation for endoscopic lesions segmentation,'' \emph{IEEE
  Trans. Circuits Syst. Video Technol.}, vol.~31, no.~5, pp. 2020--2033, 2021.

\bibitem{li2020enhanced}
M.~X. Li, Y.~M. Zhai, Y.~W. Luo, P.~F. Ge, and C.~X. Ren, ``Enhanced transport
  distance for unsupervised domain adaptation,'' in \emph{Proc. CVPR}, 2020,
  pp. 13\,936--13\,944.

\bibitem{ren2020learning}
C.~X. Ren, P.~F. Ge, P.~Y. Yang, and S.~Yan, ``Learning target-domain-specific
  classifier for partial domain adaptation,'' \emph{IEEE Trans. Neural Netw.
  Learn. Syst.}, vol.~32, no.~5, pp. 1989--2001, 2021.

\bibitem{long2015learning}
M.~Long, Y.~Cao, J.~Wang, and M.~Jordan, ``Learning transferable features with
  deep adaptation networks,'' in \emph{ICML}, vol.~37, Jul 2015, pp. 97--105.

\bibitem{ganin2016domain}
Y.~Ganin \emph{et~al.}, ``Domain-adversarial training of neural networks,''
  \emph{J. Mach. Learn. Res.}, vol.~17, no.~1, pp. 2096--2030, 2016.

\bibitem{long2018conditional}
M.~Long, Z.~Cao, J.~Wang, and M.~I. Jordan, ``Conditional adversarial domain
  adaptation,'' in \emph{Proc. NeurIPS}, 2018, pp. 1640--1650.

\bibitem{ren2019heterogeneous}
C.~X. Ren, J.~Feng, D.~Q. Dai, and S.~Yan, ``Heterogeneous domain adaptation
  via covariance structured feature translators,'' \emph{IEEE T. Cybern.},
  vol.~51, no.~4, pp. 2166--2177, 2021.

\bibitem{luo2020unsupervised}
Y.~W. Luo, C.~X. Ren, D.~Q. Dai, and H.~Yan, ``Unsupervised domain adaptation
  via discriminative manifold propagation,'' \emph{IEEE Trans. Pattern Anal.
  Mach. Intell.}, 2020, DOI: 10.1109/TPAMI.2020.3014218.

\bibitem{li2018semi}
L.~Li and Z.~Zhang, ``Semi-supervised domain adaptation by covariance
  matching,'' \emph{IEEE Trans. Pattern Anal. Mach. Intell.}, vol.~41, no.~11,
  pp. 2724--2739, 2018.

\bibitem{chen2019a}
W.-Y. Chen, Y.-C. Liu, Z.~Kira, Y.-C.~F. Wang, and J.-B. Huang, ``A closer look
  at few-shot classification,'' in \emph{ICLR}, 2019.

\bibitem{saito2019semi}
K.~Saito, D.~Kim, S.~Sclaroff, T.~Darrell, and K.~Saenko, ``Semi-supervised
  domain adaptation via minimax entropy,'' in \emph{Proc. ICCV}, 2019, pp.
  8050--8058.

\bibitem{Tseng2020Cross-Domain}
H.-Y. Tseng, H.-Y. Lee, J.-B. Huang, and M.-H. Yang, ``Cross-domain few-shot
  classification via learned feature-wise transformation,'' in \emph{ICLR},
  2020, pp. 1--16.

\bibitem{wang2021towards}
T.~Wang, Z.~Ding, W.~Shao, H.~Tang, and K.~Huang, ``Towards fair cross-domain
  adaptation via generative learning,'' in \emph{Proc. WACV}, 2021, pp.
  454--463.

\bibitem{chawla2002smote}
N.~V. Chawla, K.~W. Bowyer, L.~O. Hall, and W.~P. Kegelmeyer, ``Smote:
  synthetic minority over-sampling technique,'' \emph{J. Artif. Intell. Res.},
  vol.~16, pp. 321--357, 2002.

\bibitem{han2005borderline}
H.~Han, W.-Y. Wang, and B.-H. Mao, ``Borderline-smote: a new over-sampling
  method in imbalanced data sets learning,'' in \emph{Proc. ICIC}, 2005, pp.
  878--887.

\bibitem{he2008adasyn}
H.~He, Y.~Bai, E.~A. Garcia, and S.~Li, ``Adasyn: Adaptive synthetic sampling
  approach for imbalanced learning,'' in \emph{Jt. Conf. Neural Networks},
  2008, pp. 1322--1328.

\bibitem{batista2004study}
G.~E. Batista, R.~C. Prati, and M.~C. Monard, ``A study of the behavior of
  several methods for balancing machine learning training data,'' \emph{ACM
  SIGKDD Explor. Newsl.}, vol.~6, no.~1, pp. 20--29, 2004.

\bibitem{lin2017focal}
T.-Y. Lin, P.~Goyal, R.~Girshick, K.~He, and P.~Doll{\'a}r, ``Focal loss for
  dense object detection,'' in \emph{Proc. ICCV}, 2017, pp. 2980--2988.

\bibitem{zhou2020bbn}
B.~Zhou, Q.~Cui, X.-S. Wei, and Z.-M. Chen, ``Bbn: Bilateral-branch network
  with cumulative learning for long-tailed visual recognition,'' in \emph{Proc.
  CVPR}, 2020, pp. 9719--9728.

\bibitem{Kang2020Decoupling}
B.~Kang \emph{et~al.}, ``Decoupling representation and classifier for
  long-tailed recognition,'' in \emph{ICLR}, 2020.

\bibitem{yang2019self}
J.~Yang \emph{et~al.}, ``Self-paced balance learning for clinical skin disease
  recognition,'' \emph{IEEE Trans. Neural Netw. Learn. Syst.}, vol.~31, no.~8,
  pp. 2832--2846, 2020.

\bibitem{khalifa2020detection}
N.~E.~M. Khalifa, M.~H.~N. Taha, A.~E. Hassanien, and S.~Elghamrawy,
  ``Detection of coronavirus (covid-19) associated pneumonia based on
  generative adversarial networks and a fine-tuned deep transfer learning model
  using chest x-ray dataset,'' \emph{arXiv preprint arXiv:2004.01184}, 2020.

\bibitem{apostolopoulos2020covid}
I.~D. Apostolopoulos and T.~A. Mpesiana, ``Covid-19: automatic detection from
  x-ray images utilizing transfer learning with convolutional neural
  networks,'' \emph{Phys. Eng. Sci. Med.}, vol.~43, no.~2, pp. 635--640, 2020.

\bibitem{maghdid2021diagnosing}
H.~S. Maghdid, A.~T. Asaad, K.~Z. Ghafoor, A.~S. Sadiq, S.~Mirjalili, and M.~K.
  Khan, ``Diagnosing covid-19 pneumonia from x-ray and ct images using deep
  learning and transfer learning algorithms,'' in \emph{Multimodal Image
  Exploit. Learn.}, vol. 11734, 2021, pp. 99--110.

\bibitem{hadsell2006dimensionality}
R.~Hadsell, S.~Chopra, and Y.~LeCun, ``Dimensionality reduction by learning an
  invariant mapping,'' in \emph{Proc. CVPR}, vol.~2.\hskip 1em plus 0.5em minus
  0.4em\relax IEEE, 2006, pp. 1735--1742.

\bibitem{ren2019learning}
C.~X. Ren, P.~F. Ge, D.~Q. Dai, and H.~Yan, ``Learning kernel for conditional
  moment-matching discrepancy-based image classification,'' \emph{IEEE T.
  Cybern.}, vol.~51, no.~4, pp. 2006--2018, 2021.

\bibitem{baker1973joint}
C.~R. Baker, ``Joint measures and cross-covariance operators,'' \emph{Trans.
  Amer. Math. Soc.}, vol. 186, pp. 273--289, 1973.

\bibitem{zhang2019cpm}
C.~Zhang, Z.~Han, Y.~Cui, H.~Fu, J.~T. Zhou, and Q.~Hu, ``Cpm-nets: Cross
  partial multi-view networks,'' in \emph{Proc. NeurIPS}, vol.~32, 2019, pp.
  559--569.

\bibitem{xu2018deep}
R.~Xu, Z.~Chen, W.~Zuo, J.~Yan, and L.~Lin, ``Deep cocktail network:
  Multi-source unsupervised domain adaptation with category shift,'' in
  \emph{Proc. CVPR}, 2018, pp. 3964--3973.

\bibitem{peng2019moment}
X.~Peng, Q.~Bai, X.~Xia, Z.~Huang, K.~Saenko, and B.~Wang, ``Moment matching
  for multi-source domain adaptation,'' in \emph{Proc. ICCV}, 2019, pp.
  1406--1415.

\bibitem{kang2020contrastive}
G.~Kang, L.~Jiang, Y.~Wei, Y.~Yang, and A.~G. Hauptmann, ``Contrastive
  adaptation network for single- and multi-source domain adaptation,''
  \emph{IEEE Trans. Pattern Anal. Mach. Intell.}, pp. 1--1, 2020, DOI:
  10.1109/TPAMI.2020.3029948.

\bibitem{morozov2020mosmeddata}
S.~P. Morozov \emph{et~al.}, ``Mosmeddata: data set of 1110 chest ct scans
  performed during the covid-19 epidemic,'' \emph{Digital Diagnostics}, vol.~1,
  no.~1, pp. 49--59, 2020.

\bibitem{shi2021large}
F.~Shi \emph{et~al.}, ``Large-scale screening to distinguish between {COVID}-19
  and community-acquired pneumonia using infection size-aware classification,''
  \emph{Phys. Med. Biol.}, vol.~66, no.~6, p. 065031, 2021.

\bibitem{reddi2018convergence}
S.~J. Reddi, S.~Kale, and S.~Kumar, ``On the convergence of adam and beyond,''
  in \emph{ICLR}, 2018, pp. 1--23.

\bibitem{guan2020clinical}
W.-J. Guan \emph{et~al.}, ``Clinical characteristics of coronavirus disease
  2019 in china,'' \emph{N. Engl. J. Med.}, vol. 382, no.~18, pp. 1708--1720,
  2020.

\bibitem{maaten2008visualizing}
L.~v.~d. Maaten and G.~Hinton, ``Visualizing data using t-sne,'' \emph{J. Mach.
  Learn. Res.}, vol.~9, no. Nov, pp. 2579--2605, 2008.

\bibitem{shan2021abnormal}
F.~Shan \emph{et~al.}, ``Abnormal lung quantification in chest ct images of
  covid-19 patients with deep learning and its application to severity
  prediction,'' \emph{Med. Phys.}, vol.~48, no.~4, pp. 1633--1645, 2021.

\bibitem{si2017memory}
S.~Si, C.-J. Hsieh, and I.~S. Dhillon, ``Memory efficient kernel
  approximation,'' \emph{J. Mach. Learn. Res.}, vol.~18, no.~1, pp. 682--713,
  2017.

\end{thebibliography}

\end{document}